\documentclass[fleqn,10pt]{wlscirep}
\usepackage[utf8]{inputenc}
\usepackage[T1]{fontenc}
\usepackage{mathrsfs}

\newcommand\mnras{Mon. Not. R. Astron. Soc.}
\newcommand\apjl{Astrophys. J. Lett.}
\newcommand\apj{Astrophys. J.} 
\newcommand\aj{Astron. J.}                   
\newcommand\aap{Astron. Astrophys.}
\newcommand\araa{Annu. Rev. Astron. Astrophys.}   
\newcommand\pasp{Publ. Astron. Soc. Pac.}
\newcommand\apjs{Astrophys. J. Suppl.}     
\newcommand\ssr{Space Sci. Rev.}

\newcommand\nat{Nature}
\newcommand\prd{Physical Review D}

\title{ A seven-Earth-radius helium-burning star inside a 20.5-min detached binary }

\author[1,2,3,$\dag$]{Jie Lin}
\author[4,5,6,$\dag$]{Chengyuan Wu}
\author[7,$\dag$]{Heran Xiong}
\author[1,8,*]{Xiaofeng Wang}
\author[9,10]{P\'eter N\'emeth}
\author[4,5,6]{Zhanwen Han}
\author[4,5]{Jiangdan Li}
\author[11,12]{Nancy Elias-Rosa}
\author[11,13]{Irene Salmaso}
\author[14]{Alexei V. Filippenko}
\author[14]{Thomas G. Brink}
\author[14]{Yi Yang}
\author[4,5,6]{Xuefei Chen}
\author[1]{Shengyu Yan}
\author[4,5,6]{Jujia Zhang}
\author[15]{Sufen Guo}
\author[4,5,6]{Yongzhi Cai}
\author[1]{Jun Mo}
\author[1]{Gaobo Xi}
\author[1]{Jialian Liu}
\author[8]{Jincheng Guo}
\author[1]{Qiqi Xia}
\author[1]{Danfeng Xiang}
\author[1]{Gaici Li}
\author[4,5]{Zhenwei Li}
\author[14]{WeiKang Zheng}
\author[16,17]{Jicheng Zhang}
\author[1]{Qichun Liu}
\author[1]{Fangzhou Guo}
\author[1]{Liyang Chen}
\author[18,19]{Wenxiong Li}

\affil[1]{Physics Department and Tsinghua Center for Astrophysics, Tsinghua University, Beijing, 100084, People's Republic of China}
\affil[2]{CAS Key laboratory for Research in Galaxies and Cosmology, Department of Astronomy, University of Science and Technology of China, Hefei, 230026, People's Republic of China}
\affil[3]{School of Astronomy and Space Sciences, University of Science and Technology of China, Hefei, 230026, People's Republic of China}
\affil[4]{Yunnan Observatories, Chinese Academy of Sciences, Kunming, 650216, People's Republic of China}
\affil[5]{Key Laboratory for the Structure and Evolution of Celestial Objects, Chinese Academy of Sciences, Kunming, 650216, People's Republic of China}
\affil[6]{International Centre of Supernovae, Yunnan Key Laboratory, Kunming, 650216, People's Republic of China}
\affil[7]{
Research School of Astronomy \& Astrophysics, Australian National University, Canberra, 2611, Australia
}
\affil[8]{
Beijing Planetarium, Beijing Academy of Sciences and Technology, Beijing, 100044, People's Republic of China
}
\affil[9]{
Astronomical Institute of the Czech Academy of Sciences, Fri\v{c}ova 298, Ond\v{r}ejov, 25165, Czech Republic
}
\affil[10]{
Astroserver.org, F\H{o} t\'er 1, Malomsok, 8533, Hungary
}
\affil[11]{
INAF-Osservatorio Astronomico di Padova, Vicolo dell’Osservatorio 5, 35122 Padova, Italy
}
\affil[12]{
Institute of Space Sciences (ICE, CSIC), Campus UAB, Carrer de Can Magrans s/n, 08193 Barcelona, Spain
}
\affil[13]{
Dipartimento di Fisica e Astronomia ``G. Galilei'', Universit\`a degli Studi di Padova, Vicolo dell’Osservatorio 3, 35122 Padova, Italy
}
\affil[14]{
Department of Astronomy, University of California, Berkeley, CA 94720-3411, USA 
}
\affil[15]{
School of Physical Science and Technology, Xinjiang University, Urumqi, 830046, People's Republic of China
}
\affil[16]{Institute for Frontiers in Astronomy and Astrophysics, Beijing Normal University, 100875, People's Republic of China
}
\affil[17]{Department of Astronomy, Beijing Normal University, Beijing, 100875, People's Republic of China
}
\affil[18]{National Astronomical Observatories, Chinese Academy of Sciences, Beijing, 100101, China}

\affil[19]{The School of Physics and Astronomy, Tel Aviv University, Tel Aviv, 69978, Israel}

\affil[$\dag$]{These authors contributed equally to this work.}
\affil[*]{wang\_xf@mail.tsinghua.edu.cn}

\begin{abstract}
Binary evolution theory predicts that the second common envelope (CE) ejection can produce low-mass ($0.32-0.36\,{\rm M_\odot}$) subdwarf~B (sdB) stars inside ultrashort-orbital-period binary systems, as their helium cores are ignited under nondegenerate conditions.
With the orbital decay driven by gravitational-wave (GW) radiation, the minimum orbital periods of detached sdB binaries could be as short as $\sim 20$~minutes. However, only four sdB binaries with orbital periods below an hour have been reported so far, while none of them has an orbital period approaching the above theoretical limit. 
Here we report the discovery of a 20.5-minute-orbital-period ellipsoidal binary, TMTS~J052610.43+593445.1, 
in which the visible star is being tidally deformed by an invisible carbon-oxygen white dwarf (WD) companion.
The visible component is inferred to be an sdB star with a mass of $\sim 0.33\,{\rm M_\odot}$, approaching that of helium-ignition limit, although a He-core WD cannot be completely ruled out.
In particular, the radius of this low-mass sdB star is only 0.066~${\rm R_\odot}$, about seven Earth radii, possibly representing the most compact nondegenerate star ever known.
Such a system provides a key clue to map the binary evolution scheme from the second CE ejection to the formation of AM~CVn stars having a helium-star donor, and it will also serve as a crucial verification binary of space-borne GW detectors in the future.

\end{abstract}

\begin{document}
\flushbottom
\maketitle

\thispagestyle{empty}

\section{Main}

\begin{figure*}[htbp]
\centering
    \includegraphics[width=0.85\textwidth]{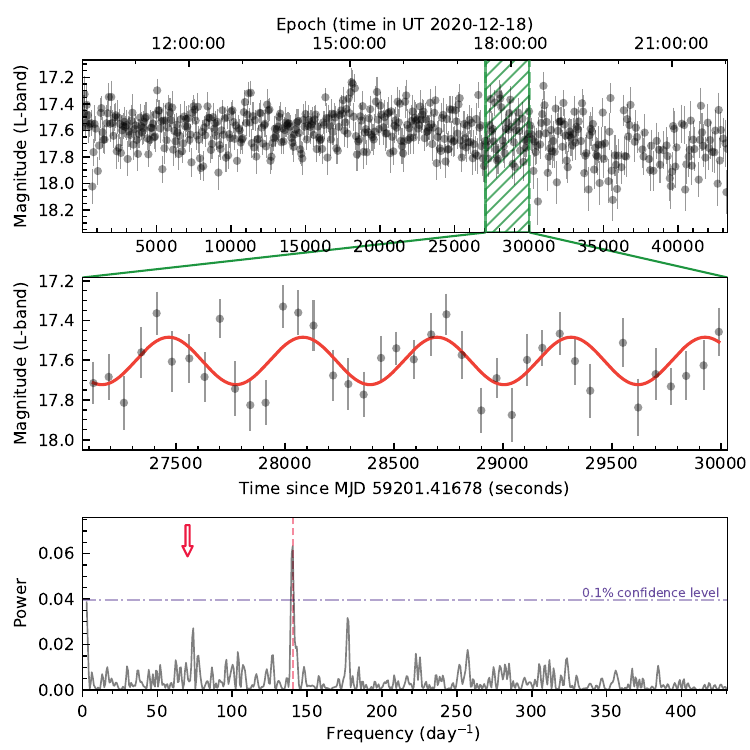}
    \caption{TMTS light curve and Lomb–Scargle periodogram for J0526.
    {\it Upper panel:} the TMTS $L$-band light curve over a 12~hr night on 18 December 2020.
    The magnitudes are presented as mean values $\pm$ 1$\sigma$.
    {\it Middle panel:} a 3000~s subset of the TMTS $L$-band light curve. The solid red line represents the best-fit sinusoidal model with a period of 10.3~min.
    {\it Lower panel:} the Lomb–Scargle periodogram (LSP) computed from the TMTS light curve. The vertical dashed line indicates the frequency corresponding to maximum power ($f_{\rm max}$). The purple dot-dashed line represents the confidence level of 0.1\%, and the red arrow shows the frequency corresponding to the orbital period (i.e., $f_{\rm max}/2$).
     } 
    \label{fig:tmts}
\end{figure*}

\begin{figure*}[htbp]
\centering
    \includegraphics[width=0.85\textwidth]{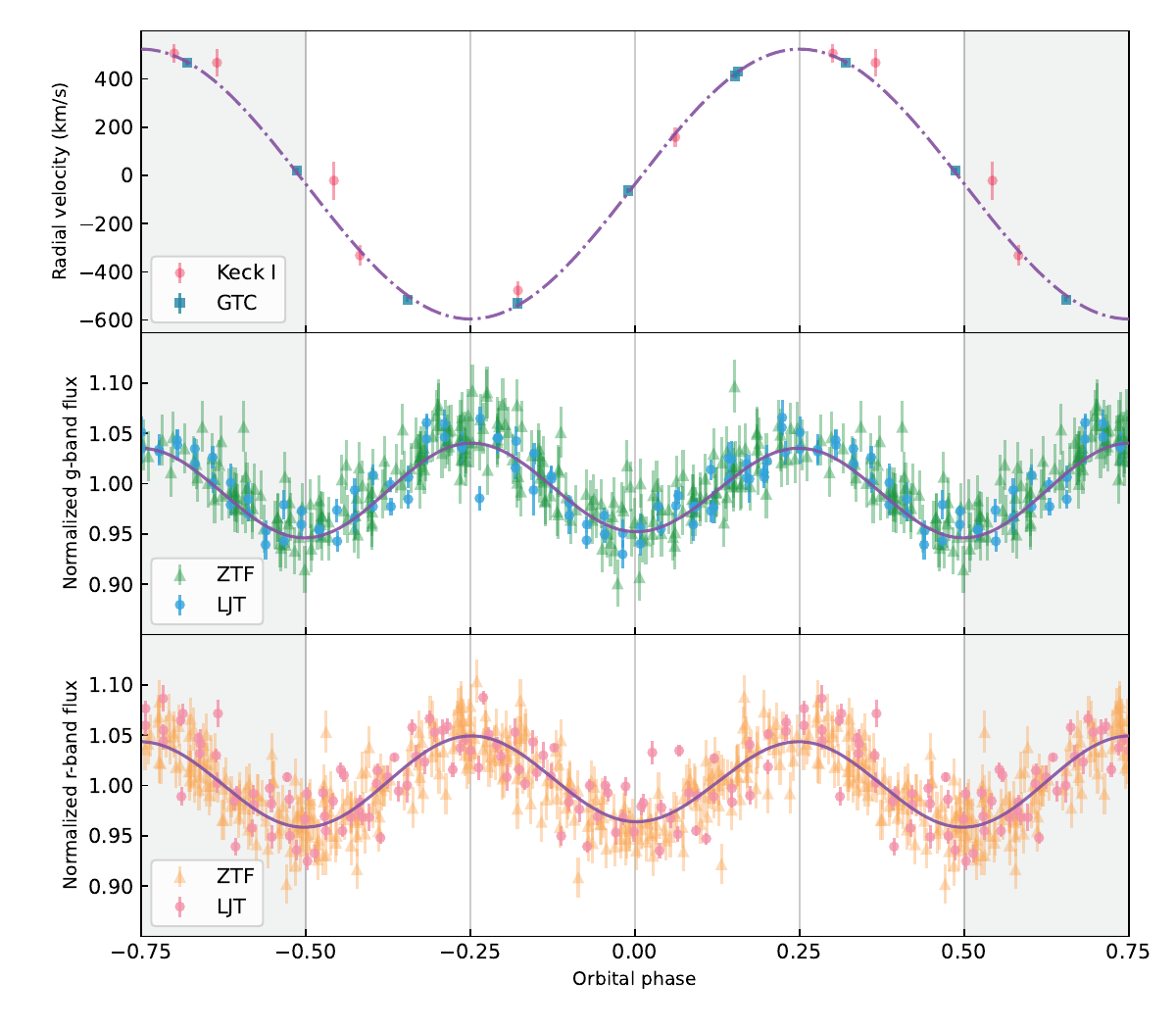}
    \caption{Phase-folded RV curve and double-band light curves for J0526.
    {\it Upper panel:} the RV curve derived from Keck/LRIS and GTC/OSIRIS observations.
    The dot-dashed line is the best-fit sinusoidal model.
    {\it Middle and lower panels:} the $g$- and $r$-band phase-folded light curves provided by LJT and ZTF.
    The purple solid lines represent the best-fit light-curve models obtained from the {\it ellc} package\cite{Maxted+2016+ellc}.
    Unequal maxima are caused by the relativistic Doppler beaming effect \cite{Loeb+Gaudi+2003+beaming_effect,Zucker+etal+2007+beaming_binary}.
    Orbital phase $\phi=0$ represents the epoch of superior conjunction when the visible star is closest to the observer. 
    The error bars represent 68\% confidence intervals throughout this paper.
     } 
    \label{fig:curves}
\end{figure*}

Since the beginning of minute-cadence observations with Tsinghua University -- Ma Huateng Telescopes for Survey (TMTS)\cite{Zhang+etal+2020+tmts_performance,Lin+etal+2021+tmts}, we have discovered a dozen unusual short-period objects \cite{lin+etal+2023+NatAs,lin+etal+2023+tmtsII} in the Galaxy.
TMTS~J052610.43+593445.1 (J2000 coordinates $\alpha = 81.5434$, $\delta = +59.5792$; hereafter J0526) is a newly discovered variable star with a dominant photometric period of only 10.3~min (see Fig.~\ref{fig:tmts}). The periodicity was cross-checked by photometric observations from the Zwicky Transient Facility (ZTF) \cite{ZTF+2019+first,ZTF+2019+products} and the Yunnan Faint Object Spectrograph and Camera (YFOSC) mounted on the Lijiang 2.4~m telescope (LJT)\cite{Lijiang+YFOSC+2015,Ljiang+performance+2019} (Fig.~\ref{fig:curves}).
Time-resolved spectroscopic observations from the Keck~I Low-Resolution Imaging Spectrometer (LRIS) \cite{Keck+LRIS+1995,Keck+LRIS+1998} and the Gran Telescope Canarias (GTC)/Optical System for Imaging and low-Resolution Integrated Spectroscopy (OSIRIS)\cite{GTC+OSIRIS+2003} yielded a dozen single-line spectra with various radial velocities (RVs; Fig.~\ref{fig:dynamical_spectra}).
The RV curve is modulated by a 20.5~min period and reaches its peaks and valleys at the phases of maximum light (Fig.~\ref{fig:curves}), which proves that J0526 is an ultracompact ellipsoidal binary.
The unequal maxima in the light curves are caused by the relativistic Doppler beaming effect \cite{Loeb+Gaudi+2003+beaming_effect,Zucker+etal+2007+beaming_binary} of the visible component, consistent with its large RV amplitude.
This object was also recently identified as a candidate verification binary (ZTF~J0526+5934) of gravitational waves (GWs) by the ZTF DR8 database and {\it Gaia} EDR3 catalog \cite{Ren+etal+2023}.

Although J0526 was included in white dwarf (WD) catalogs \cite{Gentile_Fusillo+2019+pWD,Pelisoli+Joris+2019+ELMWD}, the {\it probability of being a WD} ($P_{\rm WD}$) given by the probability map \cite{Gentile_Fusillo+2015+WD} is only $P_{\rm WD}=0.0046$ \cite{Gentile_Fusillo+2019+pWD}, suggesting that J0526 should have large differences from those WDs. 
Thus, we present a detailed analysis of J0526 in this paper. 
Since the nature of noneclipsing binaries is usually not well constrained from light curves alone, the physical parameters of J0526 were determined by the combination of spectroscopy, broad-band spectral energy distribution (SED), RV curve, and multicolor light curves.
According to the prevailing workflow for the analysis of ellipsoidal binaries, the properties of visible stars are obtained before the orbital solutions \cite{Yi+etal+2022+ns_NA,Zheng+etal+2022+ns_nearest}. Prior knowledge of the visible component helps determine the inclination of the orbital plane and the mass of the invisible component from the light curves and RV curves.
We refer to the invisible component of this binary as J0526A and the visible component as J0526B.

\subsection{Atmospheric parameters}

\begin{figure*}[htbp]
\centering
    \includegraphics[width=0.95\textwidth]{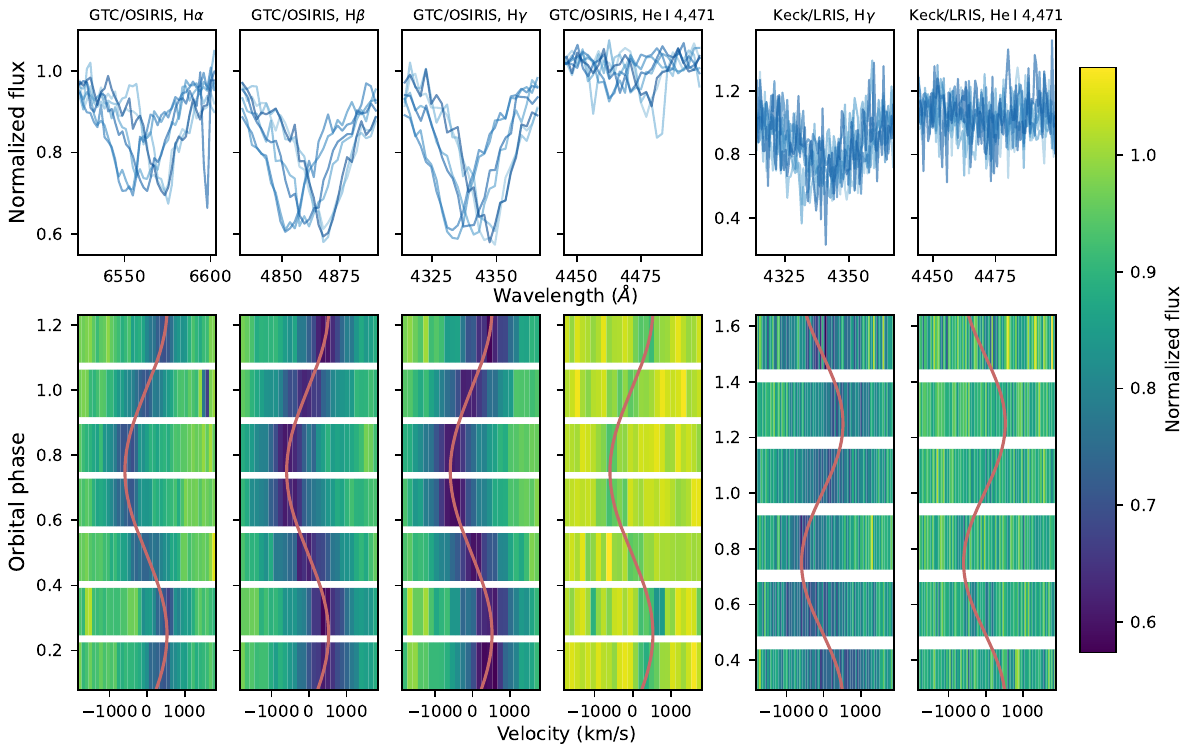}
    \caption{Dynamical spectra of J0526 from GTC/OSIRIS and Keck/LRIS observations.
    {\it Upper panels:} line profiles of H$\alpha$, H$\beta$, H$\gamma$, and He~I\ $\lambda$4471 at all epochs of observations.
    {\it Lower panels:} spectral lines phased with the orbital period of 20.5~min.
    Color scales indicate the continuum-normalized flux and red solid lines represent the best-fit RV curve of the visible star of J0526.
     } 
    \label{fig:dynamical_spectra}
\end{figure*}

As shown in the dynamical spectra (Fig.~\ref{fig:dynamical_spectra}), the Balmer lines and He~I $\lambda$4471 show synchronous shifts against the orbital phase, favoring that both H and He features arise from the visible star in the binary system. Because there are not any significant H/He lines tracing the motion of the invisible star, the invisible component must be very faint and is assumed to be negligible in the spectral fit below.
As no emission lines are visible in the spectra, mass accretion should not occur in the two components of J0526, and we thus assume that the binary system is still detached. In order to verify this assumption, we further compared the radius of the visible star with its Roche-lobe size in the following discussion.

Since the H/He absorption lines in the spectra carry key information about the atmospheric properties of the visible star, 
we fitted all of the GTC/OSIRIS spectra using non-local thermodynamic equilibrium (NLTE) spectral models obtained from  {\it TLUSTY} and {\it SYNSPEC} software \cite{hubeny17,lanz07} (see Methods).
The best-fit model reproduces well the main Balmer lines and He~I $\lambda$4471 seen in the observed spectra, which gives estimations of the effective temperature $T_{\rm eff}=25480 \pm 360\,{\rm K}$, surface gravity $\log\,g=6.355 \pm 0.068$, helium abundance $\log\,y=\log\,N_{\rm He}/N_{\rm H}=-2.305 \pm 0.062$, and projected rotational velocity $\nu\,\sin\,{i}=220 ^{+140}_{-90}\,{\rm km\,s^{-1}}$ for the visible component.

\begin{figure*}[htbp]
\centering
    \includegraphics[width=0.8\textwidth]{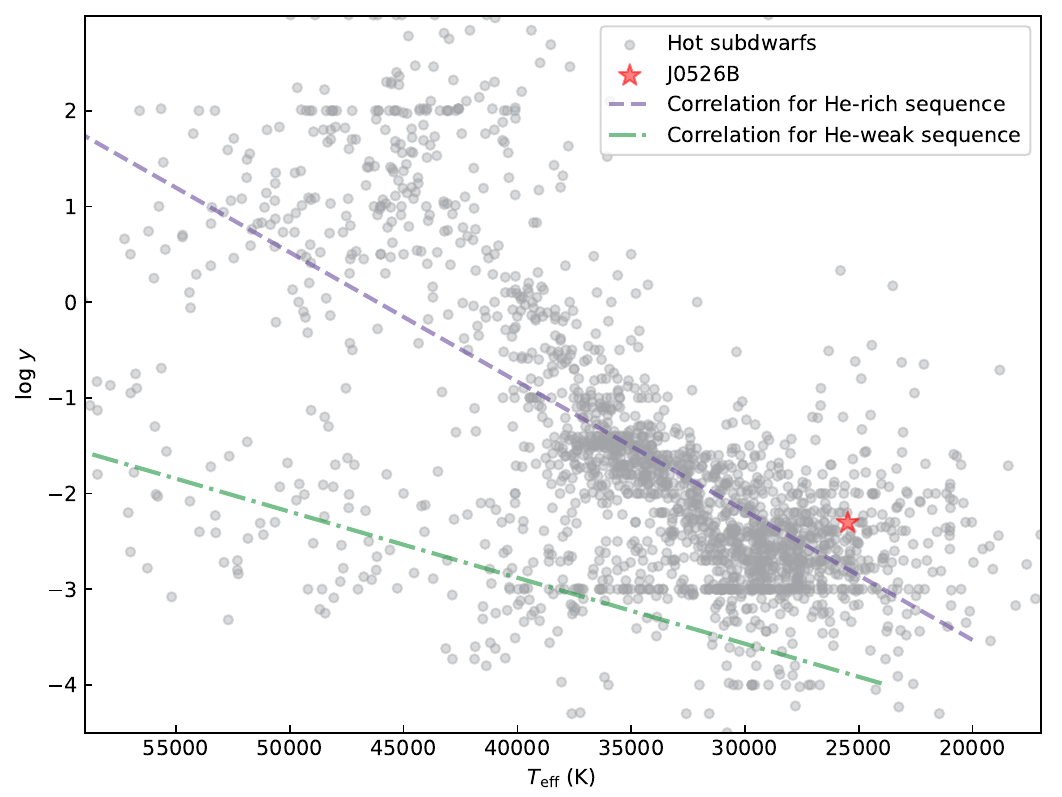}
    \caption{
    Relation between He abundance and effective temperature for hot subdwarfs. All atmospheric parameters of hot subdwarfs are taken from reference \cite{Geier+2020+sdB}. The purple dashed line and green dotted-dash line represent the correlation for the He-rich sequence \cite{Edelmann+etal+2003+sdB_cor} and the He-weak sequence \cite{Nemeth+etal+2012+sdb_cor}, respectively.
    } 
    \label{fig:abundance_temperature}
\end{figure*}

Among the large samples obtained from current spectroscopic surveys, hydrogen-rich spectra contaminated by He~I lines are very rare for WDs \cite{Gianninas+etal+2011+WD_survey+Hrich,Kepler+etal+2019+SDSS14_WD,Napiwotzki+eatl+2020+SPY_WD_survey}, but common for sdB stars \cite{Geier+2020+sdB,Lei+etal+2020+sdB,Luo+etal+2021+LAMOST_sdB}.
Observationally, the helium abundance of hot subdwarfs is overall positively correlated with their effective temperatures. However, this correlation tends to show two  distinct branches for He-rich and He-weak sequences, especially at the high-temperature end \cite{Edelmann+etal+2003+sdB_cor,Nemeth+etal+2012+sdb_cor}.
In comparison with the existing sdB sample, J0526B locates on the low-temperature end of He-rich sequence in the $T_{\rm eff}-\log\,y$ diagram  (see Fig.~\ref{fig:abundance_temperature}). 

\subsection{Stellar radius and mass}

\begin{figure*}[htbp]
\centering
    \includegraphics[width=0.6\textwidth]{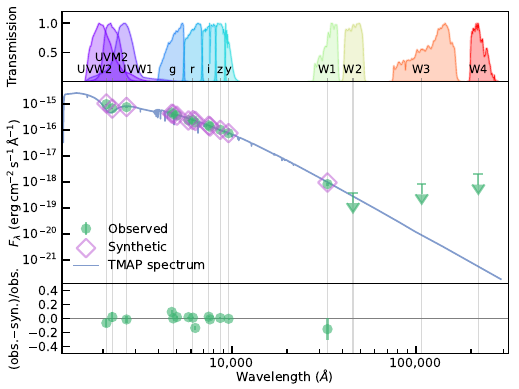}
    \caption{Broad-band SED of J0526. 
    {\it Upper panel:} transmission curves for the filters of {\it Swift}\cite{Roming+etal+2005+UVOT} ultraviolet (UV) bands, Pan-STARRS (Panoramic Survey Telescope and Rapid Response System) \cite{PanSTARRS1+2002,PanSTARRS1+2016} {\it grizy} bands, and AllWISE \cite{WISE+performance+2010,WISE+DR+2013} W1--W4 bands. All curves are normalized at the maximum.
    In order to avoid overcrowding, the transmission curves of other filters are not displayed here.
    {\it Middle panel:} the broad-band SED over UV, optical, and infrared bands.
    Green points are the photometric fluxes provided from  {\it Swift} UVW2/M2/W1 bands, {\it Gaia} DR3 \cite{Gaia_Collaboration+2016+performance,Gaia+DR3+2022} BP, G, and RP bands,  
    Pan-STARRS {\it grizy} bands, ZTF {\it gr} bands,  and AllWISE W1--W4 bands. 
    The blue solid curve represents the best-fit synthetic spectrum from the T{\"u}bingen NLTE Model-Atmosphere Package (TMAP)\cite{TMAP+soft+2012}.
    Orange diamonds indicate the synthetic fluxes derived from the model spectrum and transmission curves.
    {\it Lower panel:} relative residuals.
     } 
    \label{fig:sed}
\end{figure*}

The broad-band spectral energy distribution (SED) is a very useful tool for constraining the stellar radius and luminosity if the distance is available and reliable.
With prior knowledge of the effective temperature and surface gravity of the visible star [i.e., $T_{\rm eff,B}$ and $\log\,(g)_{\rm B}$] derived from the spectroscopic analysis above, we fit the SED (Fig.~\ref{fig:sed}) using archival multicolor photometry and the distance inferred from the {\it Gaia} DR3 \cite{Gaia+DR3+2022} parallax (see Methods).
Since this object is not included in the UV-source catalog of the Galaxy Evolution Explorer ({\it GALEX}) \cite{Galex+2005}, we also proposed UV observations with {\it Swift}/UVOT. 
The best-fit model suggests a radius $R_{\rm B}=0.0681 ^{+0.0059}_{-0.0049}\,{\rm R_\odot}$  and a bolometric luminosity $L_{\rm bol}=1.70 ^{+0.31}_{-0.24}\,{\rm L_\odot}$ for the visible star, and then updates the effective temperature and surface gravity (see Table~\ref{tab:pars}).
The model also yields an estimate of the line-sight-of extinction as $E(B-V)=0.387 ^{+0.008}_{-0.009}$ mag, well consistent with the Galactic value of $E(B-V)=0.385$ mag as queried from a three-dimensional dust extinction map\cite{Green+etal+2019+3dmap}.
With the surface gravity and radius, we computed the mass of the visible star as $M_{\rm B}=0.361^{+0.094}_{-0.073}\,{\rm M_\odot}$ using Newton's law of gravity.
The radius and mass obtained from the SED suggest that J0526B is more likely a hot subdwarf with an extremely thin hydrogen-rich envelope rather than a helium-core WD (see mass-radius relation below). 
The latter scenario is tenable only when this visible component is an inflated WD during hydrogen shell flashes, while such flashes are theoretically short-lived or even absent for the stars having such a mass
\cite{Panei+etal+2007+He_WD,Althaus+etal+2013+WDs,Istrate+etal+2016+He_core_WD}.

\begin{table}
\centering
\caption{Physical parameters of J0526. The subscript ``sd'' denotes the parameters for J0526B, and ``WD'' indicates the parameters for J0526A. The subscript ``re'' denotes those parameters refined by Bayes' theorem. 
\label{tab:pars}
}
\begin{tabular}{lc}
\hline\hline
R.A. (J2000)& $05^{\rm h}26^{\rm m}10.416^{\rm s} $\\
Dec. (J2000)& $+59^\circ 34' 45.305'' $\\
d (kpc)& $0.847 ^{+0.071}_{-0.060}$\\
$P_{\rm orb}$ (min)& $20.5062426 \pm 0.0000053$\\
$E(B-V)$ (mag)& $0.385 $\\
\hline
\textbf{Spectroscopic (GTC)}& \\
$T_{\rm eff,sd}$ (K)& $25480 \pm 360$\\
$\log(g)_{\rm sd}$ ($\rm cm\,s^{-1}$)& $6.355 \pm 0.068$\\
$\log\,(N_{\rm He}/N_{\rm H})_{\rm sd}$& $-2.305 \pm 0.062$\\
$\nu_{\rm sd}\,\sin\,{i}$ ($\rm km\,s^{-1}$)& $220 ^{+140}_{-90}$\\
\hline
\textbf{Orbital dynamics (GTC+Keck I)}& \\
$K_{\rm sd}$ ($\rm km\,s^{-1}$)& $559.6 ^{+6.4}_{-6.5}$\\
$\gamma$ ($\rm km\,s^{-1}$)& $-35.6 \pm 4.4$\\
$f(M_{\rm WD})$ ($\rm M_\odot$)& $0.259 \pm 0.009$\\
\hline
\textbf{Spectral energy distribution}& \\
$R_{\rm sd}$ ($\rm R_\odot$)& $0.0681 ^{+0.0059}_{-0.0049}$\\
$M_{\rm sd}$ ($\rm M_\odot$)& $0.361 ^{+0.094}_{-0.073}$\\
$L_{\rm bol}$ ($\rm L_\odot$)& $1.70 ^{+0.31}_{-0.24}$\\
$E(B-V)_{\rm SED}$ (mag)& $0.387 ^{+0.008}_{-0.009}$\\
$T_{\rm eff,sd,re}$ (K)& $25410 \pm 370$\\
$\log(g)_{\rm sd,re}$ ($\rm cm\,s^{-1}$)& $6.352 \pm 0.068$\\
\hline
\textbf{Light-curve analysis}& \\
$i$ (deg)& $68.2 ^{+3.7}_{-5.2}$\\
$M_{\rm WD}$ ($\rm M_\odot$)& $0.735 ^{+0.075}_{-0.069}$\\
$M_{\rm sd,re}$ ($\rm M_\odot$)& $0.360 ^{+0.080}_{-0.071}$\\
$R_{\rm sd,re}$ ($\rm R_\odot$)& $0.0661 \pm 0.0054$\\
$T_{\rm 0,re}$ ($\rm BJD_{TDB}$)& $2459933.175697 ^{+0.000016}_{-0.000017}$\\
$a$ ($\rm R_\odot$)& $0.255 \pm 0.011$\\
\hline
\textbf{Derived parameters}& \\
$(\nu_{\rm sd}\,\sin\,{i})_{\rm cal}$ ($\rm km\,s^{-1}$)& $216 ^{+18}_{-19}$\\
$\dot{P}_{\rm orb}/P_{\rm orb}$ ($\rm yr^{-1}$)& $-1.72 ^{+0.40}_{-0.47}$$\times 10^{-7}$\\
$\mathscr{A}$& $2.49 ^{+0.72}_{-0.60}$$\times 10^{-22}$\\
4-year LISA SNR& $33.6 ^{+9.7}_{-8.0}$\\
3-month LISA SNR& $3.3 ^{+1.0}_{-0.8}$\\
4-year Tianqin SNR& $24.5 ^{+7.1}_{-5.9}$\\
\hline
\end{tabular}
\end{table}

\subsection{Orbital dynamics}

The Doppler shift of spectral lines provides key clues for us to investigate the orbital dynamics of J0526.
By assuming that the orbit is circularized, the RV curve can be modeled well by a sinusoidal curve (Fig.~\ref{fig:curves}), with a RV semi-amplitude of $K_{\rm B}=559.6 ^{+6.4}_{-6.5}\,{\rm km\,s^{-1}}$ inferred for the visible component. Hence, the mass function of the invisible component was computed following
\begin{equation}
f(M_{\rm A}) \equiv \frac{M_{\rm A}^3\sin^3\,i}{(M_{\rm A}+M_{\rm B})^2} = \frac{K_{\rm B}^3P_{\rm orb} }{2\pi G} = {\rm  0.259 \pm 0.009\ M_\odot } ,
 \label{eq:mass_function}
\end{equation}
where $M_{\rm A}$ and $M_{\rm B}$ represent the masses of the invisible and visible components (respectively), $i$ is the inclination of the orbital plane, and $G$ is the gravitational constant.
Since the RV semi-amplitude and orbital period are both well-constrained from the observations, the mass function bridges a tight relation between the masses of binary and the inclination angle, and thus aids binary parameter estimation from the light-curve fit below.

\subsection{Ellipsoidal variations and compact object}

\begin{figure*}[htbp]
\centering
    \includegraphics[width=0.85\textwidth]{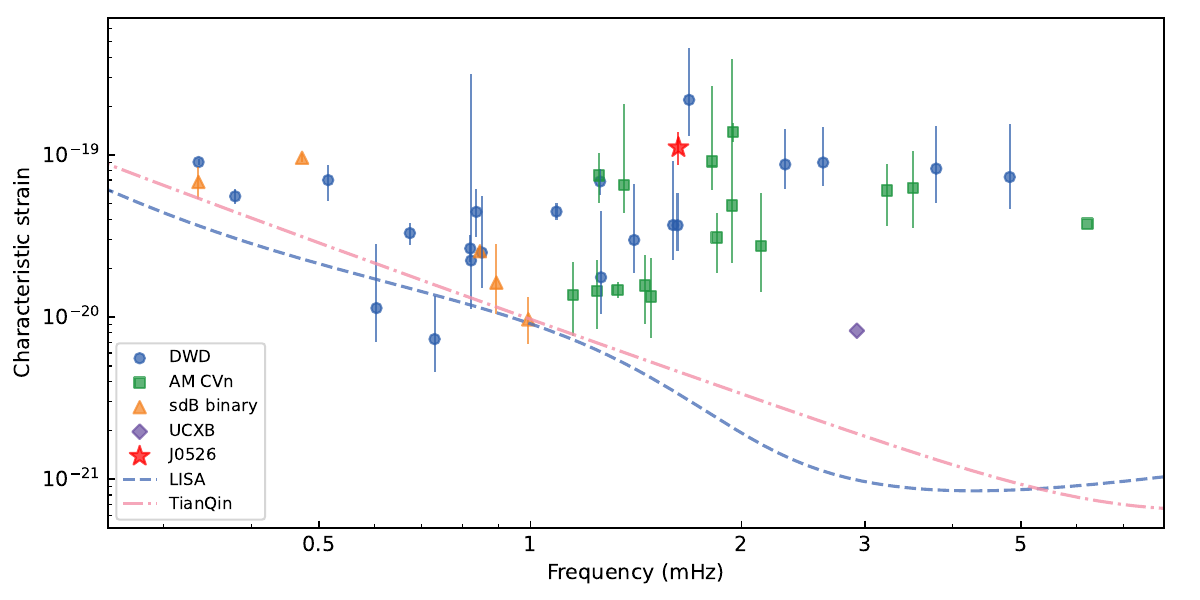}
    \caption{The characteristic strains of J0526 accompanied with dozens of verification/detectable binaries of GWs. The characteristic strains were calculated from the component masses and distances provided from reference \cite{Finch+etal+2022+GW_vbs}.
    The blue dashed line and red dotted-dash line represent the detection sensitivity curves from LISA \cite{Robson+etal+2019+LISA_sensitivity} and TianQin \cite{Huang+etal+2020+tianqin}, respectively.
    The LISA sensitivity curve here includes the instrumental noise the foreground confusion noise, while the TianQin sensitivity curve includes only the instrumental noise.
    } 
    \label{fig:GW}
\end{figure*}

Ellipsoidal modulation in the light curves is induced by tidal deformation and rotation of the visible component. Owing to the synchronization between the rotation and orbital motion, different geometric cross-sections of the visible star emerge throughout the orbit.
The large amplitude of the ellipsoidal modulations suggests that the visible star almost fills its Roche lobe (e.g., $f_{\rm R}=R_{\rm B}/R_{\rm L,B}\gtrsim 0.80$).
We modeled the $g$- and $r$-band light curves of J0526 using the {\it ellc} package \cite{Maxted+2016+ellc} (see Methods).
The values of $R_{\rm B}$, $M_{\rm B}$, and $f(M_{\rm A})$ obtained above were included as prior parameter distributions.
The Doppler beaming effect was also included in the model for offsetting the unequal maxima.
Gravity/limb-darkening coefficients and Doppler beaming factors were obtained by interpolating the grids \cite{Claret+etal+2020+LDC_WD} with the surface parameters derived from the spectroscopy and SED.
The best-fit model gives $i=68.2 ^{+3.7}_{-5.2}\,{\rm deg}$ and $M_{\rm A}=0.735 ^{+0.075}_{-0.069}\,{\rm M_\odot}$, and updates other physical parameters with Bayes' theorem (see Table~\ref{tab:pars}). 
The mass suggests that the invisible star is a carbon-oxygen (CO) WD. Through the mass-radius relation of CO WDs \cite{Parsons+etal+2017+CO_WD_MRR}, we estimate that the radius of the invisible star is about $0.011\,{\rm R_\odot}$, favoring the noneclipsing scenario for J0526.
With the updated radius of J0526B ($0.0661 \pm 0.0054\,R_{\odot}$), we derived the projected rotational velocity $(\nu_{\rm B}\,\sin\,{i})_{\rm cal}=2\pi R_{\rm B}/P_{\rm orb} =216 ^{+18}_{-19}\,{\rm km\,s^{-1}}$, consistent with the result obtained above from the spectroscopy.

Given the ultracompact orbit and relatively high masses of the two components, J0526 is predicted to be detected by Laser Interferometer Space Antenna (LISA)\cite{LISA+2017} within the first three months of its operation (see the Methods) and thus will serve as a verification binary of GWs in the future.
The GW characteristic strains of J0526 and dozens of other verification/detectable binaries of GWs are presented in Fig.~\ref{fig:GW}.

\subsection{Beyond the strip selection effects}

\begin{figure*}[htbp]
\centering
    \includegraphics[width=0.9\textwidth]{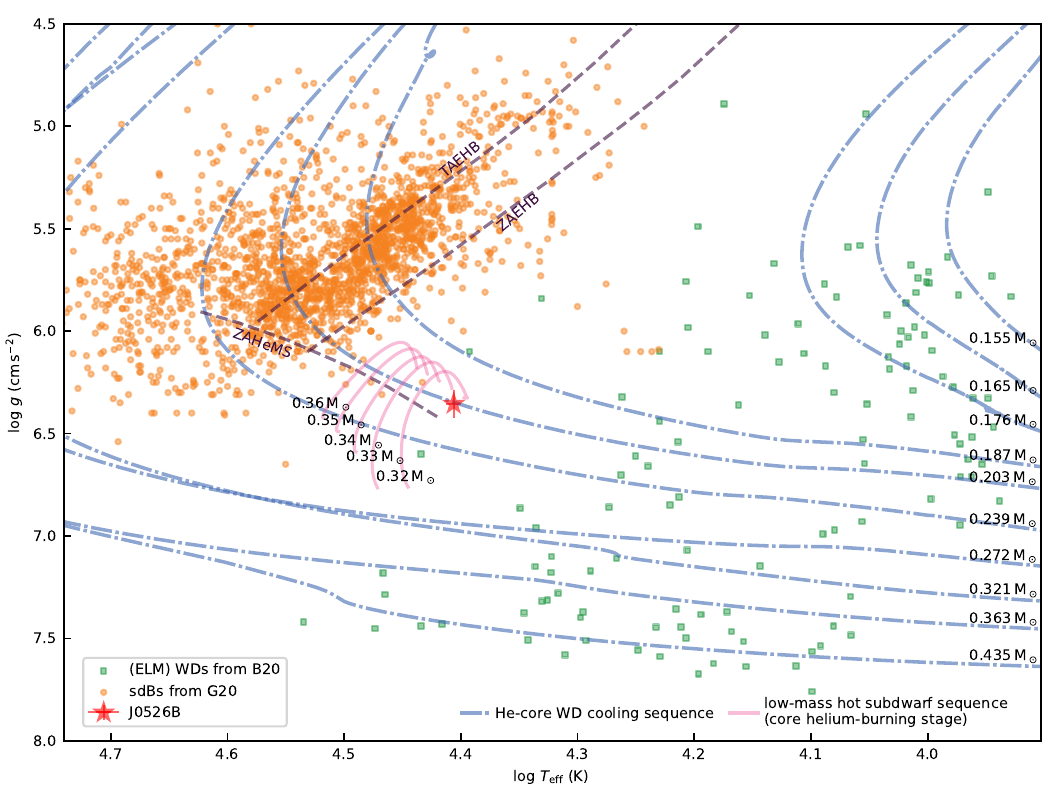}
    \caption{Kiel diagram for hot subdwarfs, low-mass WDs, and J0526B.
    The atmospheric parameters of hot subdwarfs and (extremely) low-mass WDs are taken from the references \cite{Brown+etal+2020+ELM,Geier+2020+sdB}.
    The blue dotted-dash lines are theoretical evolutionary tracks of helium-core WDs \cite{Althaus+etal+2013+WDs}, while the shell flash loops are clipped for the clarity of image.
    The purple solid lines represent the evolutionary tracks of 
    low-mass helium-core burning stars with a very thin hydrogen envelope of $10^{-6}\,{\rm M_\odot}$ (see Methods).
    We overplot the zero-age helium main sequence (ZAHeMS; see Methods), zero-age extreme horizontal branch (ZAEHB), and terminal-age extreme horizontal branch (TAEHB) \cite{Dorman+etal+1993+sdB_model} as black dashed lines.
     } 
    \label{fig:teff_logg}
\end{figure*}

Fig.~\ref{fig:teff_logg} shows that J0526B is located at an interlaced zone between hot subdwarfs \cite{Geier+2020+sdB} and (extremely) low-mass WDs \cite{Brown+etal+2020+ELM} in the Kiel diagram ($T_{\rm eff}-\log\,g$ diagram).
Since the cooling sequences of WDs are widely distributed on the blue side of the main sequence, the surface gravities and effective temperatures of hot subdwarfs are compatible with those (pre-)WDs.
We can further cross-check the nature of J0526B by comparing the spectrophotometric parallax, derived from the atmospheric parameters and hypothetical nature, against the astrometric parallax obtained from {\it Gaia} DR3 \cite{Gaia+DR3+2022}.
By assuming that J0526B is a helium-core WD, we estimated a spectrophotometric parallax $\varpi_{\rm spec}=1.542\pm0.136$~mas for J0526(see Methods), which shows a deviation from the {\it Gaia} DR3 parallax $\varpi_{\rm Gaia}=1.183\pm0.091$~mas by 2.2$\sigma$. 

Following the theoretical predictions from binary evolution theory and binary population synthesis \cite{Han+etal+2002+sdBI,Han+etal+2003+sdBII}, the second common envelope (CE) channel is responsible for the formation of sdB binaries with very short orbital periods (typically $P_{\rm orb} < 1~{\rm hr}$) \cite{Han+etal+2003+sdBII}. 
Because these sdB stars are produced from nondegenerate He cores, their masses are expected to be only $\sim 0.33\,{\rm M_\odot}$ \cite{Han+etal+2003+sdBII,Wu+etal+2018+lowmass_sdB}. 
However, the extreme horizontal branch (EHB) in Kiel diagram, corresponding to hot subdwarfs with canonical masses of $\sim 0.48\,{\rm M_\odot}$, is widely applied to confirm the natures of hot subdwarfs.
This selection effect (so-called {\it strip selection effect} \cite{Han+etal+2003+sdBII}) would lead to a systematic absence of low-mass sdB stars with very short orbits.
Additionally, for a detached sdB binary having an orbital period of only $20$~min, the hydrogen-rich envelope must be extremely thin (e.g., $10^{-6}\,{\rm M_\odot}$ \cite{Yungelson+2008+sdB_binary,Brooks+etal+2015+sdB_binary}) to avoid Roche-lobe overflow (RLOF).
With this prior knowledge, we ran the Modules for Experiments in Stellar Astrophysics ({\it MESA})\cite{Paxton+etal+2019+MESA} code to reproduce evolutionary tracks for sdB stars in extremely short-orbital-period binary systems (see Methods).
Although these tracks cover only a very small area relative to the regions of hot subdwarfs or WDs, as shown in Fig.~\ref{fig:teff_logg}, the atmospheric parameters of J0526B overlap exactly on the sdB tracks of $0.32-0.33\,{\rm M_\odot}$.
By assuming that J0526B is an sdB star of $0.33\,{\rm M_\odot}$, the spectrophotometric parallax can be re-estimated as $\varpi_{\rm spec}=1.30\pm0.10$~mas, approximating the astrometric parallax.

\subsection{Mass-radius relation}

\begin{figure*}[htbp]
\centering
    \includegraphics[width=0.95\textwidth]{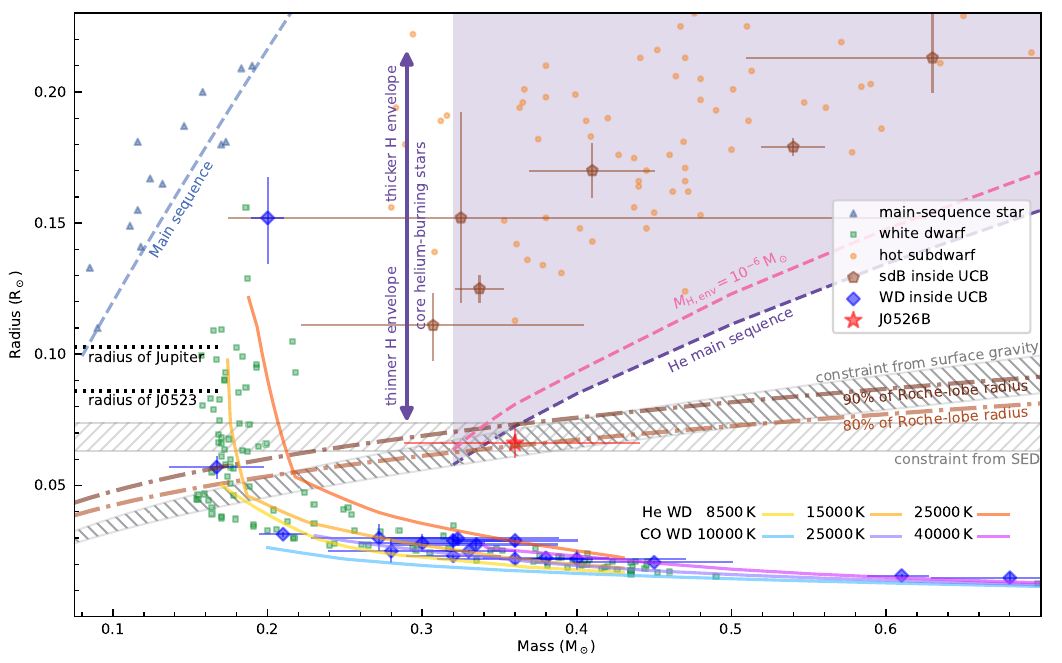}
    \caption{Mass–radius diagram for main-sequence stars, WDs, hot subdwarfs, and J0526B.
    Masses, radii of MS stars, WDs, and hot subdwarfs are collected from C17 \cite{Chen+Kipping+2017+other_worlds}, B20 \cite{Brown+etal+2020+ELM}, and S22 \cite{Schaffenroth+etal+2022+sdb_radius_mass}, respectively.
    Six sdB stars inside ultracompact binaries having orbital periods below 100~min \cite{Burdge+etal+2020+systematic_VBs,Kupfer+etal+2020+39min,Kupfer+etal+2020+56min,Geier+etal+2013+CD3011223,Pelisoli+etal+2021+sdB_NA,Kupfer+etal+2018+VBs,Finch+etal+2022+GW_vbs} are shown.
    For comparison, 18 WD components inside 11 ultracompact binaries \cite{Burdge+etal+2019+7min,Burdge+etal+2019+PTFJ0533,Burdge+etal+2020+systematic_VBs,Burdge+etal+2020+8.8min,Kupfer+etal+2020+56min} discovered by photometric surveys are also included.
    We overplot theoretical mass-radius relations for CO WDs, He~WDs, MS stars, and HeMS stars. 
    The degenerate and nondegenerate models are distinguished by solid and dashed lines, respectively.
    The purple area above the HeMS corresponds to the hot subdwarfs, i.e., the core-helium-burning (CHeB) stars with hydrogen-rich envelopes, in which thicker envelopes lead to larger stellar radii. A sequence for CHeB stars with an extremely thin hydrogen-rich envelope ($M_{\rm H,env}=10^{-6}\,{\rm M_\odot}$) is highlighted using a pink dashed line.
    We present the constraints from surface gravity and SED, derived from current observations, accompanied with the mass-radius relations of an 80\%/90\% Roche-lobe-filling star inside a 20.5~min orbital-period binary using Paczy{\'n}ski's approximation \cite{Paczynski+1971+Roche_lobe} (see Methods). 
    The radii of Jupiter and 2MASS~J0523$-$1403\cite{Dieterich+etal+2014+H_limit} (the smallest star among known MS stars) are labeled on the left of the figure.
    } 
    \label{fig:MRD}
\end{figure*}

Since diverse equations of state (EOS) from different classes of stars lead to distinct mass-radius relations, the mass–radius diagram (MRD) \cite{Ge+etal+2015+MRR,Chen+Kipping+2017+other_worlds,Lin+etal+2019+outbursts} is a valid tool for distinguishing different types of stars.
An extended version of the MRD toward the low-radius end is shown in Fig.~\ref{fig:MRD}, where the three dominant types of stars include CO-/He-core WDs, main-sequence (MS) stars, and hot subdwarfs. 
These three types of stars are located in separate regions and they hardly overlap in the MRD.
Being supported by electron degeneracy pressure, WDs tend to have smaller radii at larger masses, the opposite of nondegenerate stars.
The hot subdwarfs cover a wide area in the MRD owing to diverse hydrogen-envelope masses generated from different initial binaries and evolutionary channels.
The observational constraints from spectroscopy, the SED, and light curves support that J0526B is located exactly at the lower tip of the hot-subdwarf domain. In other words,  J0526B can be a low-mass core helium-burning (CHeB) star with an extremely thin hydrogen envelope, as also suggested by the analysis of the Kiel diagram.
As a hydrogen-exhausted star inside the 20.5~min orbit, the size of J0526B is smaller than that of all previously known nondegenerate stars, even those brown dwarfs and gas planets \cite{Dieterich+etal+2014+H_limit,Chen+Kipping+2017+other_worlds,Rappaport+eatl+2021+Hbodies}.
But J0526B has an average density $\sim 1200$ times greater than that of the Sun!

\subsection{Second CE ejection and AM~CVn stars }

Following the theoretical predictions from the second CE ejection channel of sdB stars \cite{Han+etal+2002+sdBI,Han+etal+2003+sdBII}, some sdB stars are born from a pair of WD and (evolved) MS stars, which is the stellar remnant that survived the first CE ejection \cite{Xiong+etal+2017+sdB_CE} or stable RLOF \cite{Chen+etal+2013+sdB_RLOF}.
In this channel, the WD companion has a very small radius and can penetrate deeply into the CE before CE ejection, which allows the formation of a sdB binary with a very short orbital period.
In particular, if the MS component has an initial mass larger than the critical mass for a star to experience the helium flash at the end of its first giant branch (FGB, also red-giant branch), e.g. $M_{\rm MS,0}\gtrsim 2.0\,{\rm M_\odot}$ \cite{Han+etal+2002+sdBI,Wu+etal+2018+lowmass_sdB}, 
its primary fusion reactions are through the CNO cycle instead of the proton-proton chain during its MS stage.
Consequently, the MS component can retain a higher central temperature and thus produces a more massive, nondegenerate helium core ($\sim 0.2\,{\rm M_\odot}$) when leaving the MS, compared to those with lower initial masses.
Given the higher central temperature, the helium core can be potentially ignited under nondegenerate conditions \cite{Han+etal+2002+sdBI,Chen+Han+2003} even if the envelope is lost during passage through the Hertzsprung gap.
SdB stars formed through this subchannel would have very low masses ($M_{\rm sd}=0.32-0.36\,{\rm M_\odot}$) \cite{Han+etal+2003+sdBII,Wu+etal+2018+lowmass_sdB}.
Envelopes of the Hertzsprung-gap stars are generally more tightly bound than those of stars near the tip of the FGB. 
Consequently, the inner binary system must release more orbital energy to counterbalance the binding energy, resulting in an extremely short final orbital period (typically of order tens of minutes) after the second CE ejection.

\begin{figure*}[htbp]
\centering
    \includegraphics[width=0.9\textwidth]{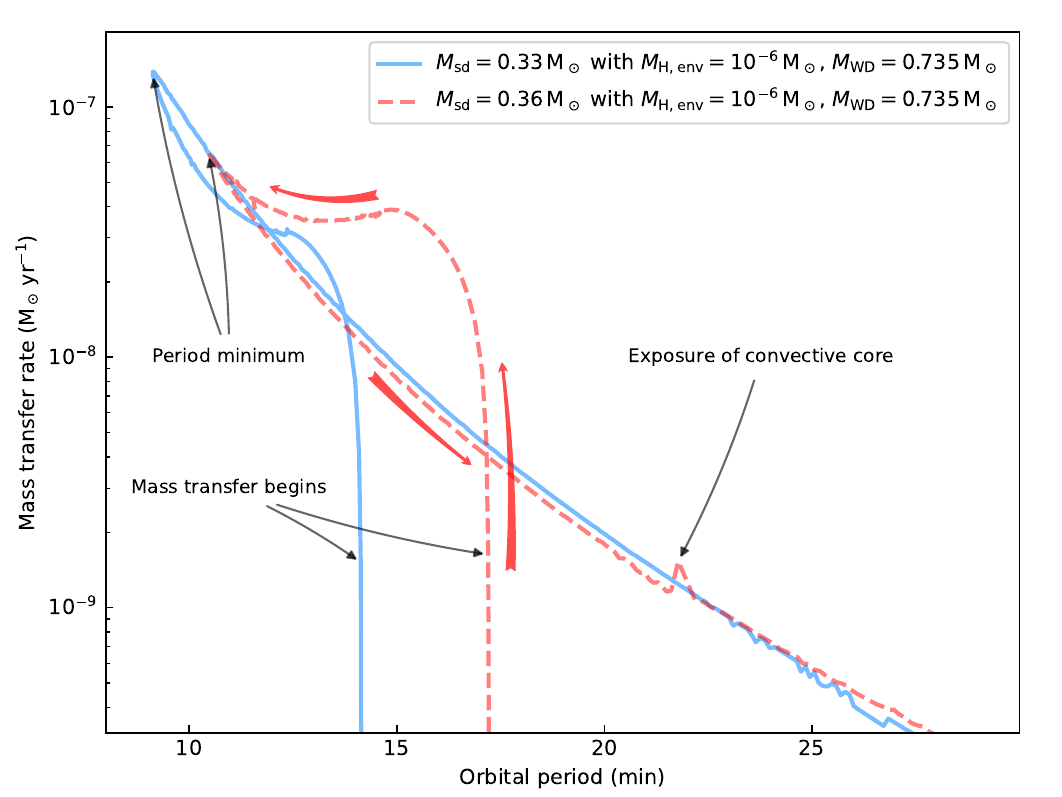}
    \caption{Binary evolution models for extremely-short-orbital-period sdB binaries.
    Two models are differentiated owing to the different core masses of sdB stars. Mass transfers are expected to begin at around 14 and 17~min for $M_{\rm sd}=0.33\,{\rm M_\odot}$ and $M_{\rm sd}=0.36\,{\rm M_\odot}$, respectively.
    The red arrows denote the direction of evolution.
    } 
    \label{fig:binary_evolution}
\end{figure*}

J0526 is such an ultracompact binary. Its extremely short orbital period, WD companion, and low-mass sdB component exactly follow the theoretical predictions from the second CE ejection channel of sdB stars \cite{Han+etal+2002+sdBI,Han+etal+2003+sdBII}.
By assuming that the mass of J0526B is equal to $0.33\,{\rm M_\odot}$ , its Roche-lobe radius is estimated as about $0.079\,{\rm R_\odot}$, which is properly wider than the size of J0526B, consistent with the above inference that this binary system is currently detached.
With orbital contraction driven by gravitational-wave radiation (GWR), after about 1.5 million years, J0526B will overflow its Roche lobe and transfer mass toward the WD at an orbital period of around 14~min (see Fig.~\ref{fig:binary_evolution}), leading to the formation of an AM~CVn star through the helium-star channel\cite{Nelemans+etal+2001+AMCVn,Solheim+2010+AMCVn}.
Owing to the nondegenerate nature of J0526B, its radius will shrink in response to mass loss induced by RLOF, supporting further orbital contraction driven by GWR.
Since the mass transfer quenches the helium burning, J0526B will begin a transition to a degenerate state, e.g. becoming a He-core WD.
When the electron-degeneracy pressure becomes dominant, J0526B will reach the minimum orbital period of $\sim 9$~min and it will start to expand with its mass loss, leading to an increasing orbital period as predicted by binary evolution theory.
Ultimately, the donor star in such an AM~CVn system either evolves into a planet orbiting the WD companion \cite{Solheim+2010+AMCVn,Blackman+etal+2021+WD_planet}, or is tidally disrupted by the WD accretor when its mass becomes smaller than  $\sim 0.002{\rm M_\odot}$ \cite{Ruderman+Shaham+1985+He_companion_TDE}.

In summary, J0526 could be the shortest-orbital-period single-degenerate detached binary, which provides crucial observational evidence supporting the complete evolutionary scheme ranging from initial binary MS stars, to MS+WD binary, to sdB+WD binary, to AM~CVn star \cite{Nelemans+etal+2001+AMCVn,Solheim+2010+AMCVn}.
With the operations of the Large Synoptic Survey Telescope (LSST)\cite{Ivezic+etal+2019+LSST}, Wide Field Survey Telescope (WFST) \cite{WFST_Collaboration+2023+WFST}, and space-borne gravitational wave observatory \cite{Luo+etal+2016+TianQin,LISA+2017}, more previously unknown extremely-short-orbital-period sdB binaries will be discovered and thus aid our understanding of the formation of sdB stars and AM~CVn stars.

\section{Methods}
\subsection{Photometric observations and orbital period}

\begin{figure*}[htbp]
\centering
    \includegraphics[width=0.9\textwidth]{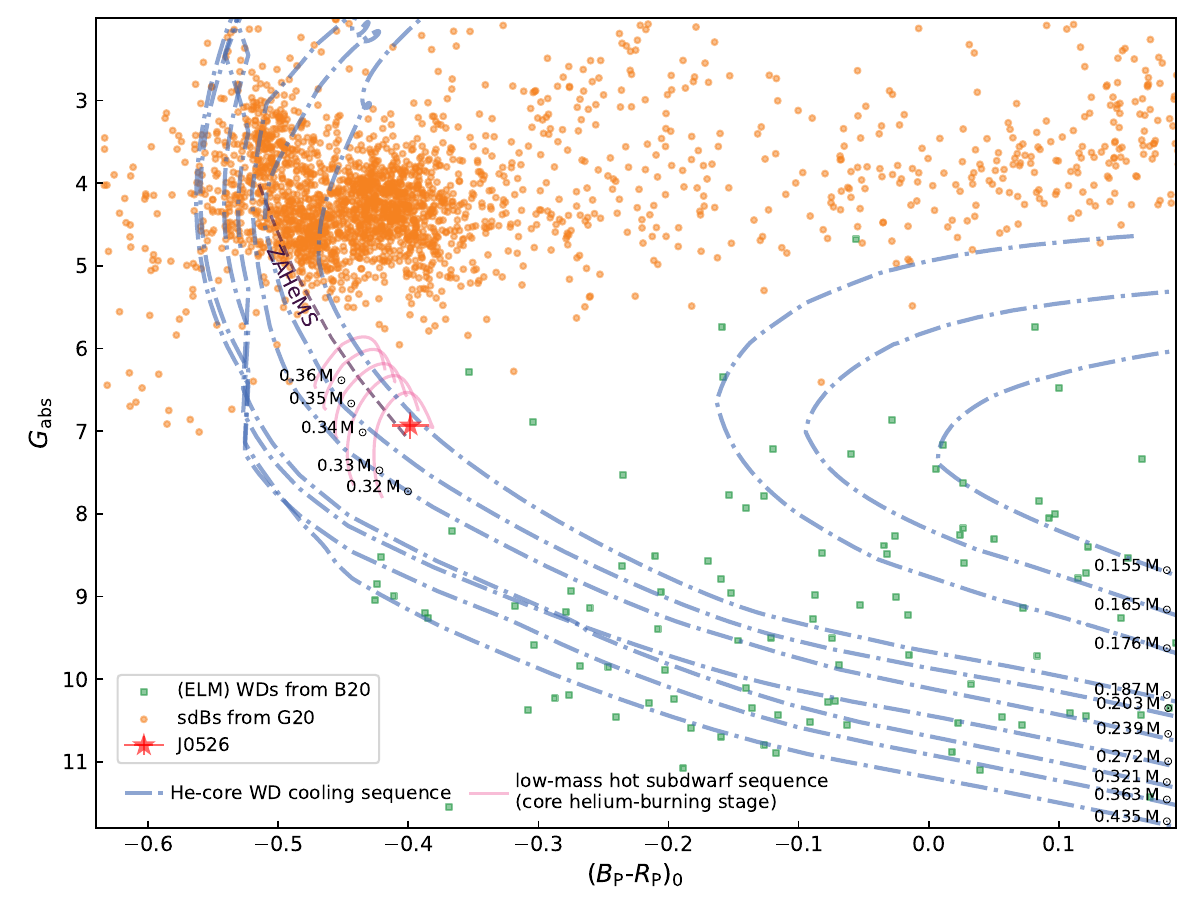}
    \caption{
    Color-magnitude diagram for hot subdwarfs, low-mass WDs, and J0526.
    The samples of hot subdwarfs and (extremely) low-mass WDs are taken from the references \cite{Brown+etal+2020+ELM,Geier+2020+sdB}, and are then cross-matched with the {\it Gaia} DR3 database\cite{Gaia+DR3+2022}. The {\it Gaia} magnitudes and colors of models are computed using the bolometric correction tables from MESA Isochrones \& Stellar Tracks (MIST)\cite{Choi+etal+2016+MIST,Dotter+2016+MIST}.
    } 
    \label{fig:hrd}
\end{figure*}

In the first two-year survey, TMTS has discovered more than 1100 variable stars with periods shorter than 2~hr \cite{lin+etal+2023+tmtsII}.
 J0526 is one of the shortest-period variable stars in the catalog.
Its 10.3~min periodicity was first revealed by the Lomb–Scargle periodogram (LSP) \cite{VanderPlas+2018+LSP_understanding} derived from the 12~hr minute-cadence observations on 18 December 2020 (UTC dates are used throughout this paper; Fig.~\ref{fig:tmts}). 
The TMTS Light-curve Analysis Pipeline automatically estimated the dereddened color $(B_{\rm p}-R_{\rm p})_0=-0.41\pm0.02$~mag and absolute magnitude $M_{\rm G}=6.86\pm0.21$ for J0526 using its embedded {\it Gaia} DR2 database \cite{Gaia_collaboration+2018+data} and {\it DUSTMAPS Python} package \cite{Green+2018+python}.
A color-magnitude diagram for J0526 with some sdB stars, low-mass WDs is presented in Fig.~\ref{fig:hrd} , using the {\it Gaia} DR3 database\cite{Gaia+DR3+2022}.
The ultrashort period and extraordinary color drove us to trigger further photometric and spectroscopic observations of this object.

The $g$- and $r$-band photometric data were obtained from ZTF Public Data Release 14 (DR14)\cite{ZTF+2019+first,ZTF+2019+products} and LJT/YFOSC observations \cite{Ljiang+performance+2019,Lijiang+YFOSC+2015} conducted on 19 December 2022.
The LJT observations lasted for 60~min in $r$ and 44~min in $g$, with a common exposure time of 30~s and  a readout time of  $\sim 3$~s.
All LJT photometric data  were reduced according to standard procedures, including bias subtraction, flat-field correction, and cosmic-ray removal.
For ZTF data, the measurements with $catflag=32768$ were excluded. 
All Modified Julian Days (MJDs) in both ZTF and LJT data were converted into $\rm BJD_{\rm TDB}$.
All observed fluxes were normalized by average fluxes in each band.

We computed the LSPs from light curves of ZTF and LJT/YFOSC observations, and thus confirmed the periodicity of J0526. Thanks to the long-term observational coverage from ZTF, we obtained a precise photometric period $P_{\rm ph}=10.2531213\pm0.0000026$~min from the $g$-band light curve, and thus the orbital period is $P_{\rm orb}=20.5062426\pm0.0000053$~min.
Since the uncertainty in the orbital period is tiny, $P_{\rm orb}$ was fixed to 20.5062426~min for the analysis below.

\subsection{Spectroscopic observations}

 We observed two series of spectra for J0526 within two independent observations runs, one with the 10~m Keck-I telescope equipped with LRIS (blue grism 600/4000,  R$\sim$1000; red grating 1200/7500, R$\sim$2000) \cite{Keck+LRIS+1995,Keck+LRIS+1998},
 and the other with the 10.4~m GTC plus OSIRIS instrument (grism R1000B, $2\times 2$ binning, R$\sim$1000 ) \cite{GTC+OSIRIS+2003}. 
 The Keck/LRIS spectra were observed at six sequential orbital phases on 23 September 2022, with an exposure time of 180~s for the first spectrum and 240~s for the others.
A total of seven GTC/OSIRIS spectra were observed on 26 January 2023, with each having an exposure time of 180~s and a readout time of $\sim 25$~s.
Because of terrible weather conditions during the Keck~I observations, the signal-to-noise ratio (SNR) of the Keck/LRIS spectra is significantly lower than that of the GTC/OSIRIS spectra.

The GTC/OSIRIS spectra were reduced following standard tasks in IRAF 
via the graphical user interface {\sc FOSCGUI},
which was designed to extract SN spectra and photometry obtained with FOSC-like instruments. It was developed by E. Cappellaro, and a package description can be found at \url{http://sngroup.oapd.inaf.it/foscgui.html}.
The raw images were first corrected for bias, overscan, trimming, and flat fielding, and subsequently one-dimensional (1D) spectra were optimally extracted from the 2D images. Wavelength calibration was performed using spectra of comparison lamps that were produced two days earlier than the observation night, while flux calibration was done via observations of spectrophotometric standard stars. These calibration images were taken with the same instrumental configuration and on the same night as the spectra of J0526. 
 Finally, the J0526 spectra were fine-tuned with the coeval broad-band photometry data, and the broad absorption bands (e.g., H$_2$O, O$_2$) due to Earth’s atmosphere were removed using the spectrum of the standard star. The Keck/LRIS spectra were reduced through a dedicated pipeline {\sl LPipe} \cite{Perley+2019+keck_pipeline} following similar procedures.

\subsection{Bayesian inference}
In order to constrain the physical parameters of J0526 from spectra, broad-band SED, RV curve,
and light curves together, we linked the model parameters derived from each observational clue using the Bayes' theorem \cite{Eadie+etal+2023+Bayesian_inference},
\begin{equation}
p( \boldsymbol{\theta}| \mathcal{D}  ) \propto \mathcal{L}(\mathcal{D}|\boldsymbol{\theta}) p(\boldsymbol{\theta}),
\label{Eq:bayesian}
\end{equation}
where $p( \boldsymbol{\theta}| \mathcal{D}  )$ is the posterior distribution of model parameters $\boldsymbol{\theta}$ with given data $\mathcal{D}$, $\mathcal{L}(\mathcal{D}|\boldsymbol{\theta})$ is the likelihood function (also sampling distribution) for $\mathcal{D}$ with given $\boldsymbol{\theta}$, and $p(\boldsymbol{\theta})$ is the prior density distribution of model parameters $\boldsymbol{\theta}$.

Following prevailing methods for resolving the nature of ellipsoidal binaries \cite{Yi+etal+2022+ns_NA,Zheng+etal+2022+ns_nearest}, we first determined the physical parameters of the visible component, which provides better parameter constraints to obtain correct orbital solutions.
The GTC spectra were fitted with non-local thermodynamic equilibrium (non-LTE) model atmospheres for obtaining 
atmospheric parameters (including $\log\,g$, $T_{\rm eff}$) of the visible component independently.
Then these atmospheric parameters were taken as prior density distributions [$p(\boldsymbol{\theta})$] to derive the radius and mass of J0526B from the broad-band SED. 
We also fitted the RV curve to obtain the RV semi-amplitude ($K_{\rm B}$) and epoch of superior conjunction ($T_0$) when the visible star was closest to the observer.
The model parameters [$\log\,(g)_{\rm B}$, $R_{\rm B}$, $K_{\rm B}$, and $T_0$ ] were used to construct prior density distributions for light-curve modeling, and were refined by final posterior distribution.

\subsection{Atmospheric model}

\begin{figure*}[htbp]
\centering
    \includegraphics[width=0.9\textwidth]{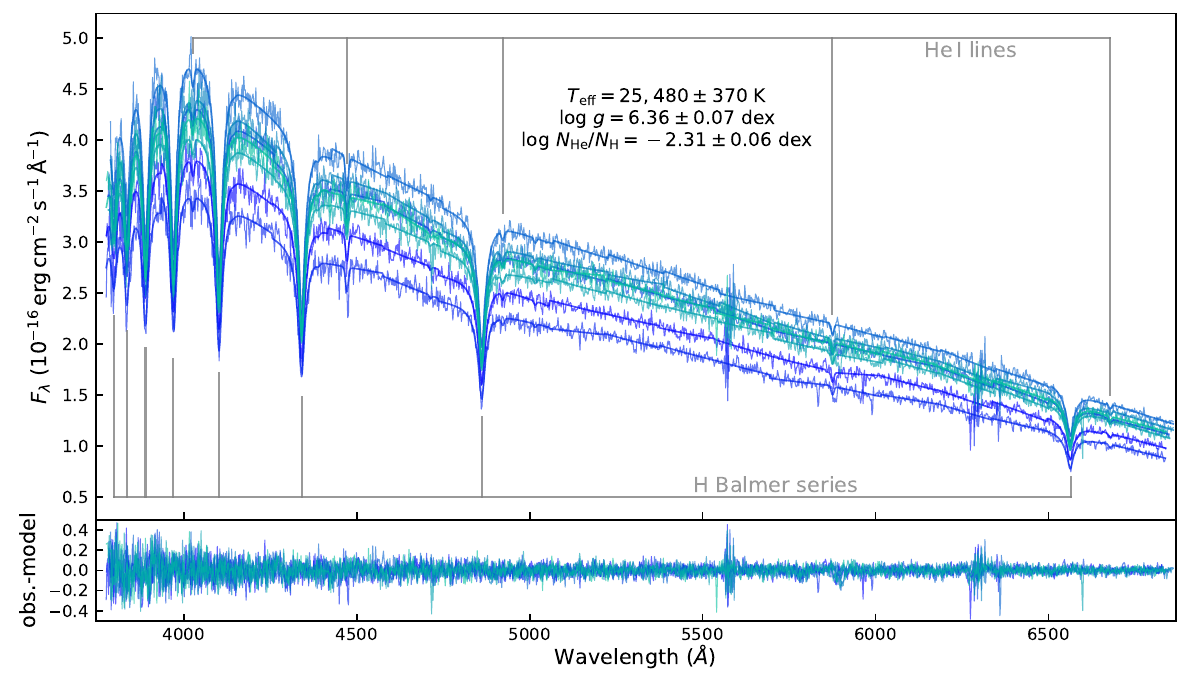}
    \caption{
   GTC/OSIRIS spectra of J0526 with their best-fit {\sc Tlusty/Synspec} synthetic spectra. 
   Residuals between observations and models are listed in the lower panel. The main H/He absorption features are indicated by grey lines.
    } 
    \label{fig:spectra_fit}
\end{figure*}

Because the SNR of the LRIS spectra is too low (owing to poor weather) to yield correct atmospheric parameters for J0526B, we adopted only the OSIRIS spectra to derive atmospheric parameters.
To reduce smearing effects and potential contributions from the invisible component, we fit all GTC spectra simultaneously (see Fig.~\ref{fig:spectra_fit}), using metal-free non-LTE {\sc Tlusty} (v207) model atmospheres and synthetic spectra with {\sc Synspec} (v53) \cite{hubeny17, lanz07}.
The best-fit model was obtained from the iterative spectral analysis procedure ({\sc XTgrid})\cite{nemeth12} which applies a steepest-descent $\chi^2$ minimization to simultaneously optimise all free parameters, including effective temperature, surface gravity, chemical abundances, and the projected rotational velocity.
All comparisons were run globally using the spectral range 3780--6850~\AA, and the model was piecewise normalized to the observations.
In parallel, the RVs were also determined by shifting each observation to the model.
The procedure converges once the relative changes of all model parameters and $\chi^2$ drop below 0.5\% in three consecutive iterations.
Finally, parameter uncertainties were calculated by mapping the parameter space around the best solution.
Notice that, when modeling the spectra of sdB stars, the systematic differences caused by replacing the metal-free non-LTE model atmospheres with the metal-line blanketed LTE model atmospheres (i.e., with [Fe/H]=$-2$ to $+1$) could be in the levels of $\Delta T_{\rm eff} \approx \pm500$~K and $\Delta \log\,g\approx \pm0.05$ \cite{Heber+2009+araa,OToole+Heber+2006,Geier+etal+2007}, which are comparable to the uncertainties quoted in our results.
All atmospheric parameters of J0526B are summarised in Table~\ref{tab:pars}.

\subsection{Spectrophotometric parallax}
A common method to test the hypothetical nature of sources is to compare their spectrophotometric parallaxes with available astrometric parallaxes.
Following the approach of calculating the spectrophotometric parallax \cite{Brown+etal+2020+ELM,Wang+etal+2022+ELM}, we first assumed J0526B to be an extremely low-mass (ELM) WD or a sdB star.
For the case of an ELM WD, we can obtain its mass $M_{\rm B}^{'}=0.237\pm0.018\,{\rm M_\odot}$ by interpolating the grids from evolutionary tracks of He-core WDs \cite{Althaus+etal+2013+WDs} using the $T_{\rm eff}$, $\log\,g$ obtained from spectra. 
The grids have included the evolutionary tracks of shell flash loops, and thus lead to  multiple solutions at same region of $T_{\rm eff}-\log\,g$ diagram.
The mass uncertainty here included the errors of surface parameters and multiple solutions of model tacks.
For the case of an sdB star, its mass is assumed to be $0.33\,{\rm M_\odot}$ based on the position of J0526B in the $T_{\rm eff}-\log\,g$ diagram (Fig.~\ref{fig:teff_logg}).
With Newton’s law of gravity and the Stefan-Boltzmann law, we estimated its stellar radius and luminosity from the atmospheric parameters and the assumed masses.
The absolute magnitude in the {\it Gaia} $G$ filter can be obtained from 
\begin{equation}
G_{\rm abs}=-2.5\log\,(\frac{L_{\rm B}}{{\rm L}_\odot})+M_{\rm bol,\odot}-BC_{G},
\label{Eq:absolute_magnitude}
\end{equation}
where $M_{\rm bol,\odot}=4.74$, and $BC_{G}$ is the bolometric correction for the $G$ band. Hence, the spectrophotometric parallax was calculated by following
\begin{equation}
G_{\rm abs}=G+5\log\,(\frac{\varpi_{\rm spec}}{100\,{\rm mas}})-A_{G},
\label{Eq:spec_parallax}
\end{equation}
where $G=17.563\pm0.003$ mag, and $A_{G}$ is the interstellar dust extinction in the $G$ band. Both
$BC_{G}$ and $A_{G}$ were obtained by interpolating the bolometric correction tables from MESA Isochrones \& Stellar Tracks (MIST)\cite{Choi+etal+2016+MIST,Dotter+2016+MIST} with the atmospheric parameters and solar metallicity.

\subsection{Radial-velocity curve}
As introduced above, the RVs were determined by shifting each observed spectrum to the {\sc Tlusty/Synspec} synthetic spectrum.
All RV measurements were corrected to the barycentric rest frame. For such a compact orbit, it is reasonable to assume that the orbit is highly circularized (i.e., eccentricity $e=0$). Therefore, the RV curve can be fitted by a sinusoidal function, and the likelihood function for RV measurements was calculated by
\begin{equation}
\ln \mathcal{L}_\nu=-\frac{1}{2}  \sum_{j} \left( \frac{ V_{{\rm r,obs,}j}- V_{{\rm r,model,}j}  }{\sigma_{{\rm \nu,obs,}j}} \right)^2,
\label{Eq:likelihood_rv}
\end{equation}
where $V_{{\rm r,model,}j} =K_{\rm B}\,\sin{[\frac{2\pi}{P_{\rm orb}} \times (t_{{\rm \nu,obs,}j}-T_0)]}-\gamma$ is the model RV at time $t_{\rm \nu,obs,j}$, and $\gamma$ is the RV of barycenter of the binary system.
Then $t_{\rm \nu,obs,j}$, $V_{\rm r,obs,j}$, and $\sigma_{\rm \nu,obs,j}$ are the BJD, observed RV, and uncertainty obtained from the $j$-th spectroscopic observation, respectively.

\subsection{Spectral Energy Distribution}

The broad-band SED was constructed from {\it Swift} \cite{Roming+etal+2005+UVOT} Target of Opportunity (ToO) observations (UV-W2/M2/W1 bands) and archival photometry from {\it Gaia} DR3 \cite{Gaia_Collaboration+2016+performance,Gaia+DR3+2022} (BP, G, and RP bands),  
    Pan-STARRS\cite{PanSTARRS1+2002,PanSTARRS1+2016} ({\it grizy} bands), ZTF \cite{ZTF+2019+first,ZTF+2019+products} ({\it gr} bands), and AllWISE\cite{WISE+performance+2010,WISE+DR+2013}  (W1--W4 bands).
The photometric measurements in three {\it Swift}-UV bands were obtained by running the {\sc HEAsoft} (ver.~6.31) command {\sc uvotproduct}. Since J0526 was accidentally located in the bad area of the detector during the observation on 15 March 2023 (ID: 00015916001), we adopted the measurements from the observations on 22 March 2023 (ID: 00015916002).
The optical and infrared photometric fluxes were directly obtained from the Virtual Observatory SED Analyzer (VOSA) \cite{Bayo+etal+2008+VOSA} online tool.
We did not use the upper-limit measurements in the WISE W2--W4 bands in the SED fitting.

We constructed the grid of synthetic photometry by sequentially integrating extinction factors and filter transmission curves over the TMAP \cite{TMAP+soft+2012}  synthetic spectra.
The extinction factors were derived from the Fitzpatrick extinction curve \cite{Fitzpatrick+1999+extinction_curve,Indebetouw+2005+extinction_curve} with reddening law $R_V=3.1$,
and the grid of reddening $E(B-V)$ spans 0.00--1.00~mag with a step of 0.05~mag. 
The transmission curves were obtained from the Filter Profile Service of Spanish Virtual Observatory (SVO) \cite{Rodrigo+Solano+2012+SVO}.
The TMAP spectrum grid covers $20,000~{\rm K}\leq T_{\rm eff} \leq 66,000~{\rm K}$ with a step of $2000$~K, and $4.50 \leq \log\,g \leq 6.50$ with a step of $0.25$. 
We applied a 3D linear interpolation to approximate the synthetic flux at arbitrary $[T_{\rm eff}, \log\,g, E(B-V)]$ coordinates within the grid.

In order to consider additional flux uncertainties caused by ellipsoidal variations of J0526, a free parameter $\sigma_{\rm F,sys}$ was included in the likelihood function,
\begin{equation}
\ln \mathcal{L}_F=-\frac{1}{2}  \sum_{j} \left\{  \frac{ [F_{{\rm obs,}j}- (R_{\rm B}\,\varpi )^2 \times F_{{\rm syn,}j}]^2 }{\sigma_{{\rm F,obs,}j}^2+\sigma_{\rm F,sys}^2}
+  \ln \left[2\pi \times (\sigma_{{\rm F,obs,}j}^2+\sigma_{\rm F,sys}^2)\right] \right\},
\label{Eq:likelihood_sed}
\end{equation}
where $F_{\rm obs,j}$ and $\sigma_{\rm F,obs,j}$ represent the observed flux and uncertainty in the $j$-th photometry, respectively.
$F_{\rm syn,j}$, the synthetic flux in the $j$-th band,  is a function of $T_{\rm eff}$, $\log\,g$ and $E(B-V)$.

\subsection{Light curves}

In order to obtain the inclination angle ( $i$) of the orbital plane and thus calculate the mass of the primary $M_{\rm A}$ through the mass function (Eq.~\ref{eq:mass_function}), we modeled all $g$- and $r$-band light curves from both ZTF and LJT observations simultaneously, using the {\it ellc} (ver~1.8.7) package \cite{Maxted+2016+ellc}.
For an ellipsoidal binary, the flux variation can be approximated as 
\begin{equation}
 \frac{\Delta F_{\rm ell}}{F} \approx 0.15\times\frac{(15+\mu_{\rm B})(1+\tau_{\rm B})}{3-\mu_{\rm B}}\frac{M_{\rm A}}{M_{\rm B}} \left (\frac{R_{\rm B}}{a} \right )^3\sin^2\,i ,
 \label{eq:ellipsoidal_modulation}
\end{equation}
where $\Delta F_{\rm ell}/F$ represents the fractional semi-amplitude of the ellipsoidal modulation, and $a$ is the orbital separation;
$\mu_{\rm B}$ and $\tau_{\rm B}$ are the Limb-darkening coefficient (LDC) and gravity-darkening coefficient (GDC) of the visible star, respectively.
Grids of LDC and GDC were generated from the tables of reference \cite{Claret+etal+2020+LDC_WD}.
We applied 2D cubic interpolations to approximate the $g$ and $r$ coefficients at arbitrary $(T_{\rm eff}, \log\,g)$ coordinates inside the grids.
With the $T_{\rm eff}$ and $\log\,g$ determined above,
we obtained
$\tau_{\rm B}=0.45$ and $\mu_{\rm B}=0.31$ for the $g$ light curves, and $\tau_{\rm B}=0.41$ and $\mu_{\rm B}=0.26$ for the $r$ band.
The Doppler beaming effect leads to higher flux  ($\phi\approx -0.25$) when the visible component is approaching the observer, than that measured in the case of moving away ($\phi\approx 0.25$). We set the beaming factor as $b=1.30$ for the $g$ band and $b=1.47$ for the $r$ band  \cite{Claret+etal+2020+LDC_WD}.

In order to respect the RV semi-amplitude $K_{\rm B}$ and epoch of superior conjunction $T_0$ obtained from the RV curves, we introduced a prior density distribution of the mass function  $f(M_{\rm B})$  (i.e., Eq.~\ref{eq:mass_function}) to bridge a tight relation among $M_{\rm A}$, $M_{\rm B}$, and inclination $i$.
In addition, the $T_0$ derived from the RV curve was also included as a prior distribution for the epoch of $\phi=0$.
Because the mass of the visible star ($M_{\rm B}$) in an ellipsoidal binary is poorly constrained by both orbital dynamics and ellipsoidal variations, we adopted the prior density distribution of $M_{\rm B}$ derived from spectroscopy and the broad-band SED.

Since the photometric data were obtained with various instruments and filters, we introduced four free parameters $\sigma_{{\rm f,sys},k}$ ($k=1,2,3,4$) for offsetting the systematic errors in ZTF $g$, ZTF $r$, LJT $g$, and LJT $r$ light curves, respectively. Hence, the likelihood function was expressed as 
\begin{equation}
\ln \mathcal{L}_f=-\frac{1}{2} \sum_{k} \sum_{j=1}^{N_k} \left\{ \frac{ (f_{{\rm obs,}j,k}-  f_{ellc,j,k})^2 }{\sigma_{{\rm f,obs,}j,k}^2+\sigma_{\rm f,sys,k}^2}
+  \ln \left[2\pi \times (\sigma_{{\rm f,obs,}j,k}^2+\sigma_{\rm f,sys,k}^2)\right] \right\},
\label{Eq:likelihood_lc}
\end{equation}
where ${N_k}$ is the number of photometry points in the $k$-th light curve, and $f_{ellc,j,k}$ represent the {\it ellc} model flux for the $k$-th band at time $t_{j,k}$. 
Then $t_{j,k}$, $f_{{\rm obs,}j,k}$, and $\sigma_{{\rm f,obs,}j,k}$ represent the BJD, observed flux, and uncertainty of $j$-th photometry point in the $k$-th light curve, respectively.

\subsection{Mass-radius relation within Roche lobe}
Given the orbital period and fillout factor (the ratio between stellar radius to its Roche-lobe radius, $f_{\rm R}=R_{\rm B}/R_{\rm L,B}$), a mass-radius relation for the star \cite{Kolb+book+2010,Lin+etal+2019+outbursts} can be obtained from Kepler's third law, 
\begin{equation}
    \frac{a^3}{P_{\rm orb}^2}=\frac{G}{4\pi^2}(M_{\rm A}+M_{\rm B})=\frac{GM_{\rm B}}{4\pi^2}\frac{1+q}{q}\, ,
    \label{eq:Kepler3}
\end{equation}
and the Paczy{\'n}ski approximation (preferred for $q=M_{\rm B}/M_{\rm A}\lesssim0.8$) \cite{Paczynski+1971+Roche_lobe} for the Roche-lobe radius (in units of orbital separation),
\begin{equation}
    f(q)=\frac{R_{\rm L,B}}{a} \approx 0.462 \left( \frac{q}{1+q} \right)^{1/3},
    \label{eq:Paczynski_approximation}
\end{equation}
where $a$ is the orbital separation. 
With $R_{\rm B}=f_{\rm R}R_{\rm L,B}=f_{\rm R}f(q)\times a$ to bridge the relation between Eq.~\ref{eq:Kepler3} and Eq.~\ref{eq:Paczynski_approximation}, the terms $q$ in the two equations cancel each other out, leading to a $q$-independent expression (if $q\lesssim0.8$ ) for the mass-radius relation,
\begin{equation}
    R_{\rm B}=0.067\,{\rm R_\odot}\times \left ( \frac{M_{\rm B}}{\rm 0.33~M_\odot } \right )^{1/3} \left ( \frac{P_{\rm orb}}{\rm 20.5~min } \right )^{2/3} \left ( \frac{f_{\rm R}}{0.85} \right ).
    \label{eq:MRR_RL}
\end{equation}

\subsection{Galactic orbit}
We computed the components of the Galactic velocity for J0526. We set the Sun at a distance of $R_0 = 8.27\pm0.29$~kpc from the Galactic center \cite{Schonrich+2012+Galactic_orbit}  and its peculiar motion in relation to the Local Standard of Rest \cite{Schonrich+etal+2010+RLS} at $(U_\odot , V_\odot , W_\odot ) = (11.1, 12.24, 7.25)~{\rm km\,s^{-1}}$. The rotational speed of the Milky Way at the Solar circle \cite{Schonrich+2012+Galactic_orbit} was set to be $V_{\rm c}= 238\pm9\,{\rm km\,s^{-1}}$. The computed Galactic velocity components resulted in ($U = 43\pm 4\,{\rm km\,s^{-1}}$, $V = 233\pm 9\,{\rm km\,s^{-1}}$, $W = 5\pm 3\,{\rm km\,s^{-1}}$), suggesting a membership in the thin-disk category \cite{Pauli+etal+2006+SPY}.

\subsection{Verification binary of gravitational waves}

For ultracompact binaries, GWs are thought to dominate the orbital angular momentum loss (AML) and thus lead to the orbital contraction. 
Thanks to the compact orbit, the GWs generated from J0526 are expected to be detected by space-borne GW observatories, such as LISA \cite{LISA+2017} and Tianqin \cite{Luo+etal+2016+TianQin}. 
With the component masses, orbital period, and distance of J0526 (Table~\ref{tab:pars}), 
we can estimate the detection SNR of its GW signal after three-month\cite{Kupfer+etal+2023+LISA} and four-year observations, respectively.

For a binary system, the GW strain amplitude can be calculated \cite{Blanchet+2014+strain} by 
\begin{equation}
    \mathscr{A}=\frac{2{(GM_c)}^{5/3}{(\pi{f}_{\rm GW})}^{2/3}}{{c}^{4}d},
\end{equation}
where $c$ is the speed of light in a vacuum, $d$ is the source distance, ${f}_{\rm GW}=2/P_{\rm orb}$ represents the GW frequency, and ${M_c}=({M}_{\rm A}{M}_{\rm B})^{3/5}/({M}_{\rm A}+{M}_{\rm B})^{1/5}$ is the chirp mass.
The characteristic strain ${h}_{\rm c}$ is thus given by ${h}_{\rm c}=\sqrt{f_{\rm GW}{T}_{\rm obs}}\mathscr{A}$,
where ${T}_{\rm obs}$ is the integration observation time, typically 4~yr for LISA and TianQin.
We present the characteristic strains for dozens of verification/detectable binaries with detection sensitivity curves from LISA \cite{Robson+etal+2019+LISA_sensitivity} and TianQin \cite{Huang+etal+2020+tianqin} in Fig.~\ref{fig:GW}.
The chirp mass of J0526 is calculated using the binary masses given by the light-curve analysis, as shown in Table~\ref{tab:pars}.
Benefiting from its close distance, J0526 is one of the ultracompact binaries that can generate the strongest GW characteristic strains.
For the first-three-month GW observation of LISA, the SNR of J0526 can reach $\sim 3$ 
(see Table~\ref{tab:pars}).

Owing to the absence of densely-sampled observations, the orbital contraction rate of J0526 is not available yet.
By assuming that the AML is driven only by GW radiation, we can obtain a theoretical 
decay rate of the orbital period using
\begin{equation}
    \frac{\dot{P}_{\rm orb}}{P_{\rm orb}}=-\frac{96}{5}\times (\frac{GM_c}{c^3})^{5/3}(\pi{{f}_{\rm GW}})^{8/3}.
\end{equation}
Hence, the theoretical period derivative of J0526 is $\dot{P}_{\rm orb}/P_{\rm orb}=-1.72 ^{+0.40}_{-0.47}\times 10^{-7} {\rm yr^{-1}} $, implying a characteristic decay timescale of $\tau_{\rm c}=\frac{3}{8}P_{\rm orb}/|\dot{P}_{\rm orb}| \approx 2.2\times10^6$~yr.

\subsection{Evolutionary models}

J0526 is an excellent object in developing gravitational-wave astronomy and investigating key process in binary evolution such as second CE ejection. In order to figure out the nature of J0526, we employed the stellar evolution code ${\sc MESA}$ \cite{Paxton+etal+2011+MESA,Paxton+etal+2013+MESA,Paxton+etal+2015+MESA,Paxton+etal+2018+MESA,Paxton+etal+2019+MESA} to investigate its origin and final fate.

\subsubsection{Single-star evolution models}

According to the observational clues introduced above, J0526 probably consists of a CO~WD and a low-mass sdB star. To construct models for the sdB star, we first create a series of He-pre-main-sequence stars, whose core masses range from $0.32\,{\rm M_\odot}$ to $0.36\,{\rm M_\odot}$ in a step of $0.01\,{\rm M_\odot}$, with He-mass fraction of $0.98$ and metallicity of $0.02$. The nuclear reaction network adopted in our simulations is ``approx21.net'' --- a $21$-isotope $\alpha$-chain nuclear network. The ``OPAL type 2'' radiative opacities tables for a carbon-/oxygen-rich mixture with a base metallicity $Z=0.02$ is employed in our simulations \cite{Iglesias+Rogers+1993+opacity_table,Iglesias+Rogers+1996+opacity_table}.
The adopted metallicity is consistent with that of Galactic thin disk membership.
Some physical processes were not included in our models such as overshooting, rotation, diffusion, etc. 

To generate the HeMS models (i.e., the CHeB models without H envelope), 
we create a series of pre-MS models with initial He abundance $0.98$ and metallicity $0.02$ (we refer to these models as ``He-pre-MS'' models) and evolve them until central-He ignition. 

In order to generate the sdB models (i.e., the CHeB models with a very thin H envelope), we evolve the ``He-pre-MS models'' until their central temperatures reach ${\rm log}{T_{\rm c}}=7.8$, and then turn off all the nuclear reactions and implant different masses of a pure H shell onto the surface of the He core using a mass accretion rate of ${10}^{-7}\,{\rm M_{\odot}\, {yr}^{-1}}$. The mass of the H shell is in the range of ${10}^{-6}$ to ${10}^{-2}\,{\rm M_\odot}$. After the formation of the H shell, we restore the nuclear reactions to evolve the sdB models forward with time.
When the nuclear reactions are restored, all the sdB models undergo Kelvin–Helmholtz contraction and thus ignite the central-He burning.
At the onset of central-He burning, their central temperatures reach ${\rm log}{T_{\rm c}}=8.0$, the central densities range from ${\rm log}{\rho _{\rm c}}=4.6$ to 4.8, and the central He mass fractions are about 0.96.
The sdB stars with core masses of $0.32-0.33\,{\rm M_\odot}$ and H-envelope mass of ${10}^{-6}\,{\rm M_\odot}$ are consistent with J0526B (see Fig.~\ref{fig:teff_logg}).

For the CHeB models with $0.32-0.33\,{\rm M_\odot}$, the stars will evolve toward the WD cooling sequence if the central He is exhausted, while other models ($0.34-0.36\,{\rm M_\odot}$) will experience unstable shell-He burning and finally become a WD when the shell He abundance cannot maintain further He burning.

\subsubsection{Binary evolutionary models}

The strong GWR from J0526 leads to orbital contraction and RLOF of J0526B in the future.
In order to predict its final fate, we evolve binary models with an initial orbital period of $20.5$~min. Among these models, the donor stars are the zero-age CHeB stars obtained above from single-star evolution. The accretor in these binaries is a $0.735~{\rm M_\odot}$ CO~WD, which is treated as a particle in the models.
To calculate the change rate of orbital angular momentum, we took into account contributions from both GWR and mass loss. The AML driven by GWR is introduced by using the standard formula \cite{Landau+Lifshitz+1975+book},
\begin{equation}
    \frac{{\rm d}\,{\rm ln}{J}_{\rm GW}}{{\rm d}t}=-\frac{32{\rm {G}^{3}}}{5{\rm {c}^{5}}}\frac{{M}_{\rm WD}{M}_{\rm sd}({M}_{\rm WD}+{M}_{\rm sd})}{{a}^{4}},
\end{equation}
where $M_{\rm WD}$ and $M_{\rm sd}$ are the masses of the WD accretor and sdB donor, respectively.

Mass transfer begins after the donor fills its Roche Lobe. However, the WD cannot accumulate all the material transferred from the donor, while some material is lost from the WD through the stellar wind, leading to the mass loss and AML of the system. 
The AML caused by the mass loss can be calculated following the expression
\begin{equation}
\dot{J}_{\rm ML,WD}=-(1-{\eta}_{\rm {He}})\dot{M}_{\rm acc}(\frac{{a}{M}_{\rm sd}}{{M}_{\rm WD}+{M}_{\rm sd}})^{2}\frac{2\pi}{{P}_{\rm orb}},
\end{equation}
where $\dot{M}_{\rm acc}$ is the mass-accretion rate of the WD, which equals the mass-transfer rate from the donor to the WD.
Referring to previous work \cite{Kato+Hachisu+2004+mass_accum,Wang+etal+2009+He_donor,Wu+etal+2021+symbiotic,Wang+etal+2017+Ia}, the mass-accumulation efficiency of WD for helium (${\eta}_{\rm {He}}$) can be approximated as
\begin{equation}
{\eta}_{\rm He}=\left\{
\begin{array}{rcl}
\frac{\dot{M}_{\rm cr, He}}{\dot{M}_{\rm He}}, & & {\dot{M}_{\rm acc}>\dot{M}_{\rm cr, He}},\\
1, & & {\dot{M}_{\rm st, He}\leq\dot{M}_{\rm acc}\leq\dot{M}_{\rm cr, He}},\\
{\eta}_{\rm He}^{'}, & & {\dot{M}_{\rm low, He}<\dot{M}_{\rm acc}<{\dot{M}_{\rm st, He}}},\\
0, & & {{\dot{M}_{\rm acc}}\leq{\dot{M}_{\rm low, He}}},
\end{array} \right.
\end{equation}
 where $\dot{M}_{\rm low, He}$ and $\dot{M}_{\rm cr, He}$ are the lowest and highest critical accretion rate (respectively), and $\dot{M}_{\rm st, He}$ is the minimum accretion rate for stable He-shell burning. If ${\dot{M}_{\rm acc}}\leq{\dot{M}_{\rm low, He}}$, all of the accreted material will be blown away by the strong stellar wind, which means that the mass-accumulation rate of the WD is almost negligible; as ${\dot{M}_{\rm low, He}<\dot{M}_{\rm acc}<{\dot{M}_{\rm st, He}}}$, the mass-accumulation efficiency under weak He-shell flashes (${\eta}_{\rm He}^{'}$) is adopted from the simulation results \cite{Wu+etal+2017+He_accretion}; 
for ${\dot{M}_{\rm st, He}\leq\dot{M}_{\rm acc}\leq\dot{M}_{\rm cr, He}}$, the He-shell burning is stable and thus all accreted material is accumulated on the surface of the WD (i.e., $\eta_{\rm He}=1$); when ${\dot{M}_{\rm acc}>\dot{M}_{\rm cr, He}}$, the WD accumulates material with its extreme rate of $\dot{M}_{\rm cr, He}$, and excess material is blown away by the optically thick wind.

The sdB stars in our models begin to transfer material to the WD after their RLOF. We evolved these models until the luminosities of the donors were below ${10}^{-6}\,{ \rm L_\odot}$. All these models experience AM~CVn phases and then the mass-transfer rates begin to decrease with increasing orbital periods, as can be seen from Fig.~\ref{fig:binary_evolution}. The final outcome is probably a WD+planet system \cite{Solheim+2010+AMCVn,Blackman+etal+2021+WD_planet} or a single WD when the donor is tidally disrupted by the accretor \cite{Ruderman+Shaham+1985+He_companion_TDE}.

\section{Data availability}
The ZTF $g$- and $r$-band photometry can be obtained from the NASA/IPAC Infrared Science Archive (\url{https://irsa.ipac.caltech.edu}).
The optical and infrared photometric fluxes in the SED can be obtained from the VOSA online tool (\url{http://svo2.cab.inta-csic.es/theory/vosa}).
The bolometric correction tables can be downloaded from MIST (\url{http://waps.cfa.harvard.edu/MIST/model_grids.html}).
All light curves, observed and synthetic spectra, RV curve, photometric and synthetic fluxes in SED,  stellar/binary evolutionary models used for this work are available at our Zenodo page ( \url{https://www.zenodo.org/record/8074854} or \url{https://doi.org/10.5281/zenodo.8074854}) .

\section{Code availability}
The codes of {\sc Tlusty} (v207) and {\sc Synspec} (v53) that are used for generating (non-LTE) model atmospheres and producing synthetic spectra are available at \url{https://www.as.arizona.edu/~hubeny}, and the services of online spectral analyses ({\sc XTgrid}) are provided from Astroserver (\url{www.Astroserver.org}).
The {\sc python} package {\it ellc} (v1.8.7) for modeling light curves can be obtained from \url{https://pypi.org/project/ellc}.
The sensitivity curve of LISA can be computed using the codes from \url{https://github.com/eXtremeGravityInstitute/LISA_Sensitivity}.
The software {\sc MESA} (v12778) used for stellar evolutionary calculations is available at \url{http://mesastar.org}, and the full inlists for evolutionary models used for this work are available at our Zenodo page ( \url{https://www.zenodo.org/record/8074854} or \url{https://doi.org/10.5281/zenodo.8074854}) .

\section*{Acknowledgements}

We acknowledge the support of the staffs of the 10.4~m Gran Telescopio Canarias (GTC), Keck~I 10~m telescope, Lijiang 2.4~m telescope (LJT), and {\it Swift}/UVOT.
The work of X.-F.W. is supported by the National Natural Science Foundation of China (NSFC grants 12033003, 12288102, and 11633002), the Ma Huateng Foundation, the New Cornerstone Science Foundation through the XPLORER PRIZE, China Manned-Spaced Project (CMS-CSST-2021-A12), and the Scholar Program of Beijing Academy of Science and Technology (DZ:BS202002). 
J.L. is supported by the Cyrus Chun Ying Tang Foundations.
C.-Y.W. is supported by the National Natural Science Foundation of China (NSFC grant 12003013) and the Yunnan Fundamental Research Projects (No. 202301AU070039).
C.-Y.W., Z.-W.H., X.-F.C., J.-J.Z. and Y.-Z.C. are supported by International Centre of Supernovae, Yunnan Key Laboratory (No. 202302AN360001).
P.N. acknowledges support from the Grant Agency of the Czech Republic (GA\v{C}R 22-34467S). 
The Astronomical Institute in Ond\v{r}ejov is supported by the project RVO:67985815. 
N.E.R. acknowledges partial support from MIUR, PRIN 2017 (grant 20179ZF5KS) ``The new frontier of the Multi-Messenger Astrophysics: follow-up of electromagnetic transient counterparts of gravitational wave sources.'', from PRIN-INAF 2022 ``Shedding light on the nature of gap transients: from the observations to the models'', from the Spanish MICINN grant PID2019-108709GB-I00 and FEDER funds.
I.S. is supported by funding from MIUR, PRIN 2017 (grant 20179ZF5KS), and PRIN-INAF 2022 project ``Shedding light on the nature of gap transients: from the observations to the models'', and acknowledges the support of the doctoral grant funded by Istituto Nazionale di Astrofisica via the University of Padova and the Italian Ministry of Education, University and Research (MIUR).
A.V.F.'s group at U.C. Berkeley has received financial assistance from the Christopher R. Redlich Fund, Alan Eustace (W.Z. is a Eustace Specialist in Astronomy), Briggs and Kathleen Wood (T.G.B. is a Wood Specialist in Astronomy), Gary and Cynthia Bengier, Clark and Sharon Winslow, and Sanford Robertson (Y.Y. is a Bengier-Winslow-Robertson Postdoctoral Fellow), and many other donors.
Y.-Z. Cai is supported by the National Natural Science Foundation of China (NSFC, Grant No. 12303054).

This research is based on observations made with the Gran Telescopio Canarias (GTC), installed at the Spanish Observatorio del Roque de los Muchachos of the Instituto de Astrof\'{\i}sica de Canarias, on the island of La Palma. This research is based on data obtained with the instrument OSIRIS, built by a Consortium led by the Instituto de Astrof\'{\i}sica de Canarias in collaboration with the Instituto de Astronom\'{\i}a of the Universidad Nacional Aut\'onoma de Mexico. OSIRIS was funded by GRANTECAN and the National Plan of Astronomy and Astrophysics of the Spanish Government.
Some of the data presented herein were obtained at the W. M. Keck Observatory, which is operated as a scientific partnership among the California Institute of Technology, the University of California, and the National Aeronautics and Space Administration  (NASA); the observatory was made possible by the generous financial support of the W. M. Keck Foundation. 
We acknowledge the Target of Opportunity (ToO) observations supported from {\it Swift} Mission Operations Center.
This research has used the services of \mbox{\url{www.Astroserver.org}} under reference T4JRRH and Y75AKG.

Based in part on observations obtained with the Samuel Oschin 48-inch Telescope at the Palomar Observatory as part of the Zwicky Transient Facility project. ZTF is supported by the U.S. National Science Foundation (NSF) under grant AST-1440341 and a collaboration including Caltech, IPAC, the Weizmann Institute for Science, the Oskar Klein Center at Stockholm University, the University of Maryland, the University of Washington, Deutsches Elektronen-Synchrotron and Humboldt University, Los Alamos National Laboratories, the TANGO Consortium of Taiwan, the University of Wisconsin at Milwaukee, and Lawrence Berkeley National Laboratories. Operations are conducted by COO, IPAC, and UW.

This work has made use of data from the European Space Agency (ESA) mission {\it Gaia} (\url{https://www.cosmos.esa.int/gaia}), processed by the {\it Gaia} Data Processing and Analysis Consortium (DPAC, \url{https://www.cosmos.esa.int/web/gaia/dpac/consortium}). Funding for the DPAC has been provided by national institutions, in particular the institutions participating in the {\it Gaia} Multilateral Agreement.

The Pan-STARRS1 Surveys (PS1) and the PS1 public science archive have been made possible through contributions by the Institute for Astronomy, the University of Hawaii, the Pan-STARRS Project Office, the Max-Planck Society and its participating institutes, the Max Planck Institute for Astronomy, Heidelberg and the Max Planck Institute for Extraterrestrial Physics, Garching, The Johns Hopkins University, Durham University, the University of Edinburgh, the Queen's University Belfast, the Harvard-Smithsonian Center for Astrophysics, the Las Cumbres Observatory Global Telescope Network Incorporated, the National Central University of Taiwan, the Space Telescope Science Institute, NASA under grant NNX08AR22G issued through the Planetary Science Division of the NASA Science Mission Directorate, NSF grant AST-1238877, the University of Maryland, Eotvos Lorand University (ELTE), the Los Alamos National Laboratory, and the Gordon and Betty Moore Foundation.

This publication makes use of data products from the Wide-field Infrared Survey Explorer, which is a joint project of the University of California, Los Angeles, and the Jet Propulsion Laboratory/California Institute of Technology, funded by NASA.

This publication makes use of VOSA, developed under the Spanish Virtual Observatory (https://svo.cab.inta-csic.es) project funded by MCIN/AEI/10.13039/501100011033/ through grant PID2020-112949GB-I00. 
VOSA has been partially updated by using funding from the European Union's Horizon 2020 Research and Innovation Programme, under Grant Agreement no 776403 (EXOPLANETS-A).

\section*{Author contributions statement}
J.L., C.-Y.W., H.-R.X. and X.-F.W. drafted the manuscript; Z.-W. H. and A.V.F. edited the manuscript in detail; P.N., N.E.R., X.-F.C., Y.-Z.C. and S.-F.G. also helped with the manuscript. 
X.-F.W. is the PI of TMTS and led the discussions.
J. L. discovered this source by analysing the large-volume data from TMTS observations and performed detail analysis in spectroscopy, SED, orbital dynamic and light curves.
C.-Y.W. computed the stellar/binary evolution models for the low-mass sdB stars, and H.-R.X. provided some key ideas for these models. 
P.N. determined the atmospheric parameters from GTC/ OSIRIS spectra, and computed radial velocities from both GTC/OSIRIS and Keck-I/LRIS spectra.
J.-D.L. and Q.-Q.X. helped with the analysis of SED and light curves.
The GTC/OSIRIS spectra were provided and reduced by N.E.R. and I.S..
A.V.F., T.G.B., Y.Y. and W-K.Z. obtained and reduced the Keck-I data.
J.-J.Z. obtained and reduced the high-cadence observations of the Lijiang 2.4\,m telescope.
S.-F.G. computed the Galactic orbit.
J.-L.L. reduced and analyzed the observations of {\it Swift}/UVOT.
S.-Y.Y., Y.-Z.C., J.-C.G., D.-F.X. and G.-C.L. assisted in the spectral observations and analysis.
J.L., C.-Y.W., H.-R.X., X.-F.W., P.N., Z.-W. H., J.-D.L., X.-F.C., J.-C.G., Q.-Q.X. and Z.-W.L. contributed to beneficial discussions.
X.-F.W., J.-C.Z., J.M., G.-B.X. and J.L. contributed to the building, pipeline, and database of TMTS.
G.-B.X., J.M., J.-C.G., Q.-Q.X., Q.-C.L., F.-Z.G., L.-Y.C. and W.-X.L. contributed to the operations and data products of TMTS.


\end{document}